\documentclass[aps,prb,twocolumn,superscriptaddress]{revtex4}
\usepackage{graphicx, bm, amsmath}
\usepackage{microtype}
\usepackage{upgreek}
\usepackage{amssymb}

\begin{document}

\title{Emergence of Quantum Phase-Slip Behaviour in Superconducting NbN Nanowires: DC Electrical Transport and Fabrication Technologies} 

\author{N.G.N. Constantino}
\affiliation{London Centre for Nanotechnology, University College London, 17-19 Gordon Street, London, WC1H 0AH, United Kingdom}
\author{M.S. Anwar}
\affiliation{London Centre for Nanotechnology, University College London, 17-19 Gordon Street, London, WC1H 0AH, United Kingdom}
\author{O.W. Kennedy}
\affiliation{London Centre for Nanotechnology, University College London, 17-19 Gordon Street, London, WC1H 0AH, United Kingdom}
\author{M. Dang}
\affiliation{London Centre for Nanotechnology, University College London, 17-19 Gordon Street, London, WC1H 0AH, United Kingdom}
\author{P.A. Warburton}
\affiliation{London Centre for Nanotechnology, University College London, 17-19 Gordon Street, London, WC1H 0AH, United Kingdom}
\author{J.C. Fenton} 
\affiliation{London Centre for Nanotechnology, University College London, 17-19 Gordon Street, London, WC1H 0AH, United Kingdom}
\email{j.fenton@ucl.ac.uk}

\begin{abstract}
Superconducting nanowires undergoing quantum phase-slips have potential for impact in electronic devices, with a high-accuracy quantum current standard among a possible toolbox of novel components. A key element of developing such technologies is to understand the requirements for, and control the production of, superconducting nanowires that undergo coherent quantum phase-slips. We present three fabrication technologies, based on using electron-beam lithography or neon focussed ion-beam lithography, for defining narrow superconducting nanowires, and have used these to create nanowires in niobium nitride with widths in the range of 20--250 nm. We present characterisation of the nanowires using DC electrical transport at temperatures down to 300 mK. We demonstrate that a range of different behaviours may be obtained in different nanowires, including bulk-like superconducting properties with critical-current features, the observation of phase-slip centres and the observation of zero conductance below a critical voltage, characteristic of coherent quantum phase-slips. We observe critical voltages up to 5 mV, an order of magnitude larger than other reports to date. The different prominence of quantum phase-slip effects in the various nanowires may be understood as arising from the differing importance of quantum fluctuations. Control of the nanowire properties will pave the way for routine fabrication of coherent quantum phase-slip nanowire devices for technology applications.

~

This article has been published as {\em Nanomaterials} \textbf{2018}, {\em 8}, 442. \copyright{The authors}. The article is distributed under a Creative Commons (CC-BY) licence.
\end{abstract}

\maketitle

\section{Introduction}
Superconducting nanowires have attracted interest for several decades, both fundamental interest in superconductivity in reduced dimensions and for the development of applications. Applications in technological devices have grown out of these studies, with photon detectors, relevant to application in both astronomy \cite{doylereview} and quantum technologies \cite{hadfieldreview}, being probably the most significant application to date. Interest in a new range of devices in superconducting nanowires was sparked by the publication in 2006 of a proposal \cite{mooijnazarov} that superconducting nanowire devices could form a new class of component elements, based on a phenomenon termed coherent quantum phase-slip (CQPS). In such devices~\cite{hriscu}, the superconducting nanowire operates as the charge--flux dual device to the Josephson junction, a device that itself led to the development of many technologically useful devices following its discovery in the 1960s. The charge--flux duality means that the roles of quantum conjugate variables charge (or charge number) and magnetic flux (or superconducting phase) are exchanged. (This also implies the exchange of other related quantities including current-voltage and inductance-capacitance.) Particular interest has been attracted by the prospect that the superconducting nanowires could form the basis of a quantum current standard, the charge--flux dual to the Josephson voltage standard. In such a device, the time-averaged transport current passing along a voltage-biased nanowire as it is irradiated with microwaves exhibits step-like structures as the voltage bias is varied, and these could be used to calibrate the current, potentially very precisely. 

Quantum phase-slips may occur in superconducting nanowires with cross-sectional dimensions $\lesssim\xi$, where $\xi$ is the superconducting coherence length, the shortest characteristic length on which the properties of a superconductor can vary. A quantum phase-slip (QPS) process may actually be coherent or incoherent. Coherent quantum phase-slips involve coherent quantum tunnelling between states corresponding to different numbers of flux quanta on the two sides of the nanowire, such~that the system may end up in superpositions of these states. This is the charge--flux dual of the Josephson junction, in which the behaviour results from coherent quantum tunnelling between states corresponding to different numbers of Cooper pairs on the two sides of a weak link. 
If, on the other hand, a QPS process is not sufficiently strong to lead to a final state that is a superposition of states corresponding to different numbers of flux quanta on the two sides of the nanowire, incoherent quantum phase-slips may still occur. These comprise individual 2$\pi$ phase-slips occurring within the nanowire, with associated dissipation, which shows up as a measurable resistance for the nanowire. Each phase-slip event corresponds to the transfer of one quantum of magnetic flux from one side of the nanowire to the other. Incoherent QPS are conceptually related to (also incoherent) thermally activated phase-slips (TAPS) \cite{tinkham}, where incoherent QPS involves tunnelling through the energy barrier between the states and TAPS involves thermal activation over the energy barrier. However, it is {coherent} quantum phase-slips that are of use for charge--flux-dual applications.

The characteristic macroscopic manifestation of coherent QPS in the DC geometry is zero conductance below a critical voltage $V_\textrm{c}$. Although this has similarities to Coulomb blockade features in Josephson junctions and tunnel junctions \cite{chandrasekhar,haviland}, strong evidence that coherent quantum phase-slip effects are indeed to be found in suitably fabricated superconducting nanowires has been provided through several careful experiments \cite{astafiev,websterfenton,hongisto,peltonen,degraaf}, including a recent report \cite{degraaf} in which a gate tuned the interference between two CQPS elements in the expected way. A number of these convincing experiments were RF spectroscopy experiments \cite{astafiev,peltonen,degraaf} and were carried out in a geometry that does not allow characterisation of the QPS nanowire by DC transport, the geometry relevant for the proposed quantum current standard. To date, there have been very few reports in DC transport of {{coherent}} QPS effects. The basic signature of CQPS in DC transport, a current blockade below a critical voltage, has been reported in a NbSi nanowire by Webster \textit{et al.} \cite{websterfenton} with $V_{\textrm{c}} \approx 500$ $\upmu$V. \mbox{Hongisto \textit{et al.} \cite{hongisto}} also reported $V_\textrm{c}$ features in two NbSi nanowires, up to 480 $\upmu$V, and furthermore showed evidence for quantum interference between two CQPS elements. Critical voltage features were also observed in \cite{arutyunovjsnm,kafanov}.

In DC transport measurements on devices with related physics, measurements of the temperature dependence of resistance, $R(T)$, have historically been very useful for their characterisation. Measurements of a residual resistance remaining in superconducting nanowires just below the superconducting transition temperature $T_\textrm{c}$ in the 1970s revealed the presence of TAPS. In later studies of the superconductor--insulator transition in both thin films \cite{goldman,paalanen} and nanowires \cite{bollinger,lau}, measurements of samples' $R(T)$ enabled characterisation of the samples' properties, showing a dependence of $T_\textrm{c}$ on the disorder level, with $\textrm{d}R/\textrm{d}T$ for $T>T_\textrm{c}$ giving an indication of the proximity to the superconductor--insulator transition. Both early measurements on superconducting nanowires in a DC-transport configuration relating to observation of QPS, and many reports since, have likewise focussed on the $R(T)$ behaviour. Non-zero resistance persisting below $T_\textrm{c}$ at temperatures lower than expected for TAPS has been observed by multiple groups, and this has frequently been interpreted as evidence for incoherent QPS. Although the underlying effect in incoherent QPS is quantum tunnelling, a temperature dependence nonetheless arises as a result of the temperature dependence of other physical properties of the system, particularly the superconducting coherence length $\xi$. Physical models of QPS, which have adjustable parameters, have been successfully fitted to such $R(T)$ data \cite{arutyunov}.
Since~there are other possible explanations of a persisting residual resistance below $T_\textrm{c}$, further tests, such as fitting to the current-voltage dependence $I(V)$ below $T_\textrm{c}$, are valuable. While in fact very few of the past reports have included such information on the $I(V)$ dependence, \mbox{Altomare \textit{et al.} \cite{altomare}} and \mbox{Makise \textit{et al.} \cite{makise}} reported low-temperature $I(V)$ matching expectations for QPS behaviour, strengthening, for these measurements, if not more widely, the interpretation that the tail in the $R(T)$ arises from incoherent~QPS.

The absence of further reports of the $V_\textrm{c}$ feature that is characteristic of CQPS in DC $I(V)$ measurements, despite, anecdotally, experimental efforts by several research groups, hints at experimental challenges in realising all the requirements for CQPS in the technologically useful DC-transport geometry \cite{fentonjpcs}. QPS are expected in nanowires with cross-sectional dimensions $\sim\xi$, and another key requirement is maximising the characteristic energy scale for quantum phase-slips, $E_\textrm{S}$. In~a nanowire with cross-sectional dimensions $\leq\xi$, $E_\textrm{S}$ may be expressed \cite{mooijnjp} as:

\begin{equation} \label{eq}
E_\textrm{S}=a\frac{l}{\xi}k_\textrm{B}T_\textrm{c}\frac{R_\textrm{Q}}{R_\xi}\exp{\bigg(-b\frac{R_\textrm{Q}}{R_\xi}\bigg)}
\end{equation}
where $l$ is the length of the nanowire, $R_\xi=R_{\Box}\xi/w$ is the normal-state resistance of a coherence length of nanowire, $R_{\Box}=R_\textrm{N}w/l$ its sheet resistance, $R_\textrm{N}$ the low-temperature normal-state resistance, $w$~the nanowire width, $R_\textrm{Q}=h/(4e^2)=6.45$ k$\Omega$ the resistance quantum for Cooper pairs and $a$ and $b$ numerical constants of order unity. $R_\xi$ should therefore be maximised in order to maximise $E_\textrm{S}$.
Coherent QPS should only be expected in the limit $E_\textrm{S}\gg k_\textrm{B}T$. The characteristic voltage scale for the critical voltage in a simple CQPS nanowire is $V_\textrm{c}=2\pi E_\textrm{S}/(2e)$, and $V_\textrm{c}$ is also the characteristic scale for the width of voltage steps at constant current in the proposed dual-Shapiro effect, which provides another motivation for maximising $E_\textrm{S}$. 

Mooij \textit{et al.} \cite{mooijnjp} also showed that the environment a nanowire is embedded in affects whether or not quantum fluctuations in a CQPS nanowire lead to a blockade of current. Specifically, they argued that a current blockade occurs ({i.e.}, CQPS drive the nanowire into an insulating state) for values of the ratio $E_\textrm{S}/E_L$ greater than a critical value $\alpha_\textrm{c}$, where $E_L=\Phi_0^2/(2L)$ is the inductive energy associated with the series inductance $L$ in the circuit and $\Phi_0=h/(2e)$ is the quantum of magnetic flux. A larger series inductance in the circuit therefore promotes the formation of a current blockade.
Additionally, as for the Josephson voltage standard \cite{kautz}, in order to obtain stable step features, for the charge--flux dual current standard, there are requirements on the parameters of the embedding circuit. This means that a series resistance of an appropriate value should be included in a CQPS-nanowire current-standard circuit \cite{fentonjpcs}.

As shown by the form of Equation \ref{eq}, for $E_\textrm{S}$ to be non-negligible, the superconducting material must have a large value of $R_\xi$. This requires a high low-temperature sheet-resistance $R_{\Box}$, implying a high kinetic inductance. Obtaining homogeneous materials satisfying this requirement is arguably more demanding than satisfying the requirement that the cross-sectional dimensions of the nanowire be of the order of $\xi$. To satisfy the requirement of large $R_{\Box}$, convenient materials systems to work with for CQPS applications are two-component superconductors, which exhibit a superconductor--insulator transition as the composition varies. Close to the superconductor--insulator transition, on the superconducting side of it, the material exhibits a high normal-state resistivity, as required for CQPS devices. 
The $T_\textrm{c}$ of the nanowires must also be high enough that the device can be operated far below it, so as to minimise the existence of thermally excited quasiparticles, although in practice, to date, this condition has been less stringent than the requirement for coherent quantum phase-slips that $E_\textrm{S}\gg kT$. In the Bardeen--Cooper--Schrieffer (BCS)~theory of superconductivity, the coherence length scales inversely with $T_\textrm{c}$, so~there is a trade-off between $\xi$ and $T_\textrm{c}$; a system with workable values for both parameters must be chosen. Several candidate material systems have been used for superconducting nanowires, including InO$_x$, NbN, NbSi, MoGe and Ti \cite{fentonjpcs,kim,astafiev,degraaf,peltonen,websterfenton,hongisto,makise,lau,arutyunovjsnm,arutyunovscirep,kafanov,manninen}.

The detailed properties of materials at the nanoscale are often dependent on the means of fabrication, and so it is important that several means of generating nanowires are assessed. Experimental realisations of these superconducting nanowires have focussed either on the use of molecular templating techniques \cite{lau,bollinger} or on the use of subtractive fabrication techniques \cite{arutyunov,kim}. In the latter, a thin film of the superconducting material is first deposited onto a substrate and then processed to remove material and leave a narrow nanowire. For fabricating nanowires with widths below 50 nm, lithography using an electron beam or a focussed ion beam is a natural choice for defining the nanowires. Deposition onto clean substrates followed by removal of unwanted superconducting-film material by etching is preferred to lift-off-based fabrication as it avoids both issues of film contamination by the resist during film deposition and the need to use a double-layer resist to avoid so-called `lily padding' of the deposited film where metal deposited on sidewalls remains attached to the sides of the nanowire following lift-off.

As already indicated, the nanowires should be embedded in a high-impedance environment. This also isolates the nanowire from environmental influence. Thin-film inductors may be formed by wider lengths of the same superconducting material, which have substantial kinetic inductance in the superconducting state, but which are wide enough that their quantum phase-slip rate is negligible. If~thin-film resistors are required, these may be formed by depositing an additional material to form part of the circuit \cite{nash}.

In practice, controlling the properties of the superconducting material close to the superconductor--insulator transition is far from trivial. The superconducting properties of a material are modified in very thin films and also in very narrow nanowires within such films. Inhomogeneities are present as a result of randomness in the film deposition process. Such inhomogeneities both may become accentuated during the etching process and also become more relevant to the electrical properties in a narrow nanowire as the width decreases towards the length scale of inhomogeneities in the film. Furthermore, close enough to the superconductor--insulator transition, an inhomogeneous electronic state of the film is expected on theoretical grounds \cite{feigelman} to be induced even in a perfect material by very small variations due to offset charges in the substrate. 

In this article, we report investigations carried out in niobium nitride. We have investigated the variation in the properties of the material as its dimensions are reduced down to nanowires. We assess three different fabrication technologies for creating narrow nanowires and demonstrate their use for generating nanowires with width dimensions below 20 nm. We report a range of superconducting and CQPS properties that we have observed in such nanowires, interpret the results, discuss their implications and outline challenges that will be relevant to others seeking to carry out similar studies.

\section{Results}
In this section, we introduce the nanowires fabricated by the three different fabrication technologies we have employed. We report both investigations of the changes in properties of the niobium nitride films as the thickness is varied, revealed by $R(T)$ measurements, and then report representative $I(V)$ behaviour of a number of niobium nitride nanowires, showing a range of behaviours, including the emergence of behaviour characteristic of coherent quantum phase-slips.

\subsection{Film Characterisation: Changes from Bulk Properties to Thin-Film Behaviour}
\begin{figure*}[!ht]
\begin{center}
\includegraphics[width=5 cm]{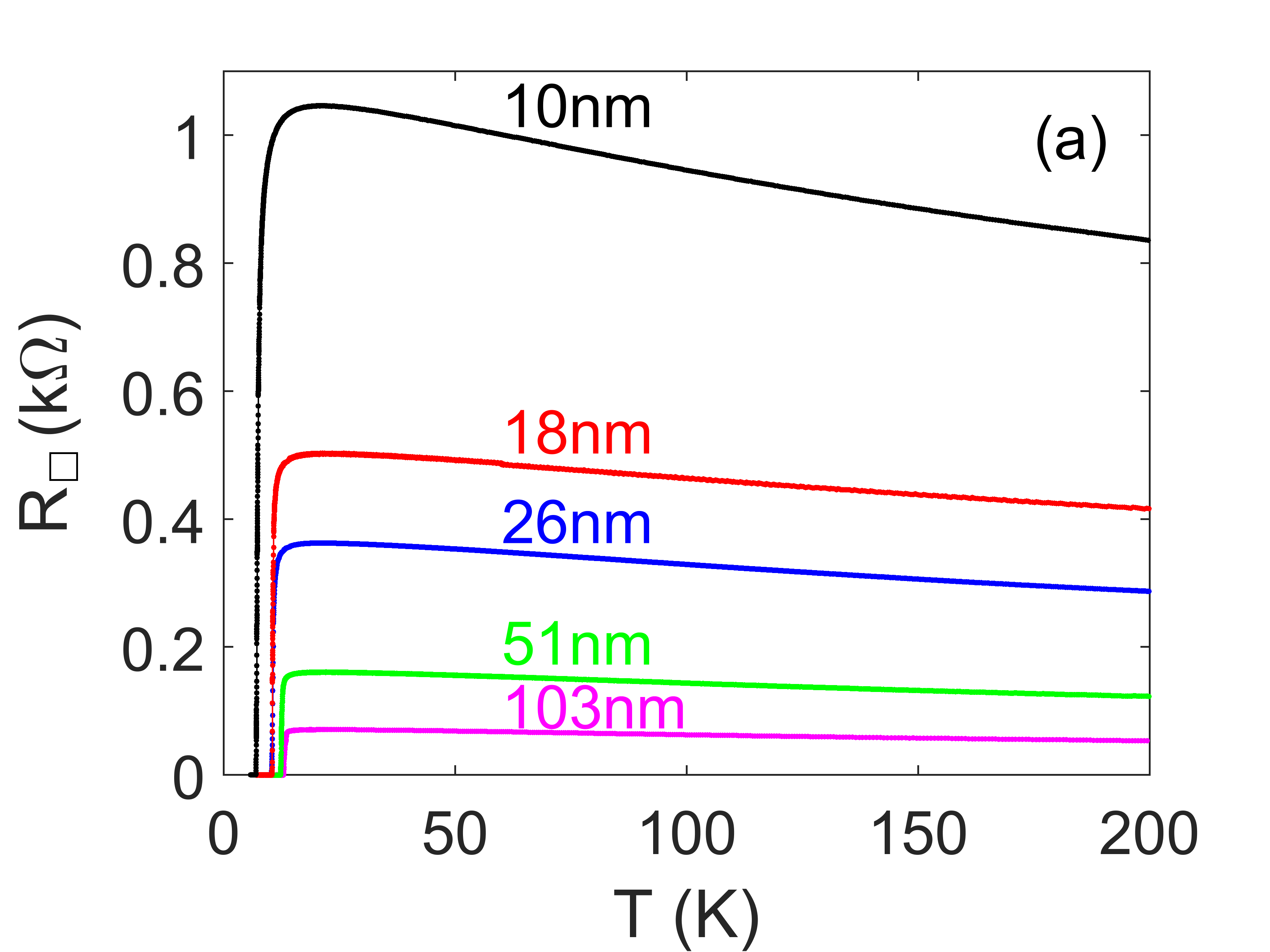}
\includegraphics[width=5 cm]{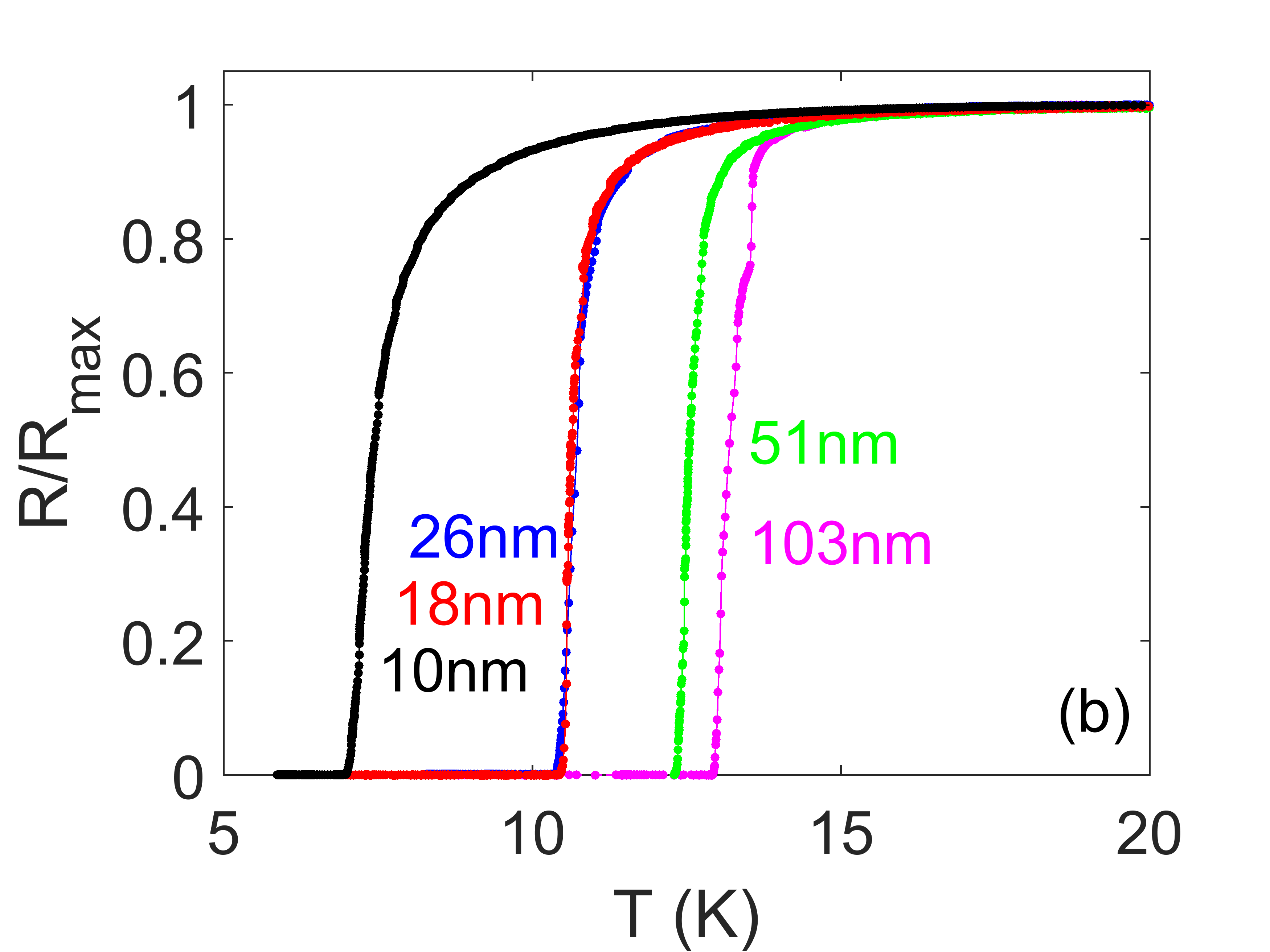}
\includegraphics[width=5 cm]{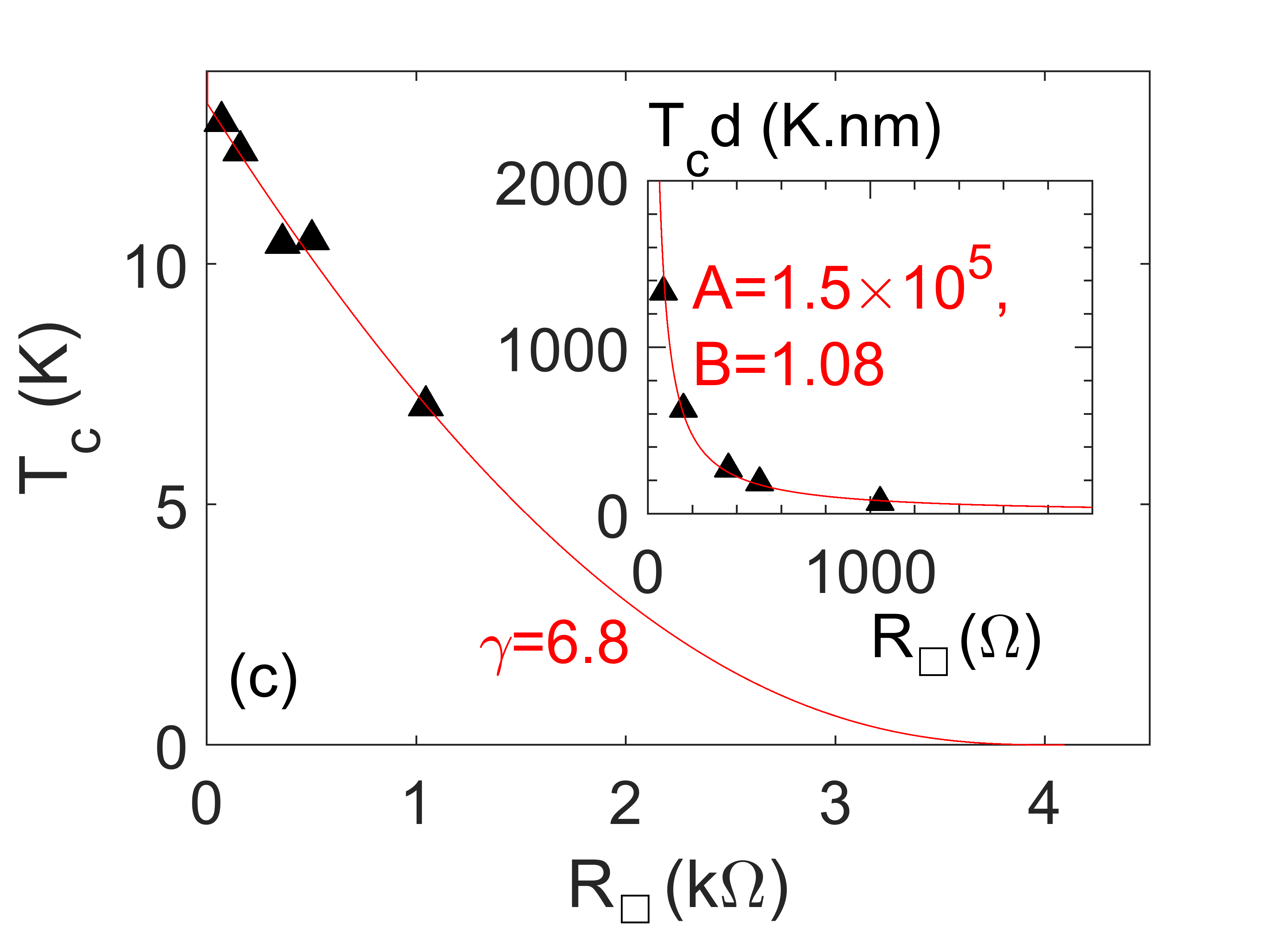}
\caption{\label{NCfilms} Measurements and analysis of $R(T)$ for NbN films with thickness $d$ in the range 10--103~nm. (\textbf{a}) Sheet resistance $R_{\Box}(T)$. The films with thickness of 10 nm and 26 nm were patterned into structures with 90-$\upmu$m-wide, 2.4-mm-long tracks prior to measurement; the other measurements were carried out on unpatterned films. (\textbf{b}) The same data as (a) at low $T$, normalised to the maximum resistance $R_\textrm{max}$. (\textbf{c})~Variation of $T_\textrm{c}$ with low-temperature maximum $R_{\Box}$. The line shows a fit to $T_\textrm{c}/T_\textrm{c0}=\exp{(\gamma)}((1/\gamma-\sqrt{t/2}+t/4)/(1/\gamma+\sqrt{t/2}+t/4))^{1/\gamma}$ with $t=R_{\Box}/(4\pi R_\textrm{Q})$ and where the fit coefficient $\gamma$ is related to the elastic scattering time $\tau$ by $\tau=(h/(k_\textrm{B}T_\textrm{c0}))\exp{(-\gamma)}$ \cite{finkelstein}, and we also treat $T_\textrm{c0}$ as a fit coefficient, obtaining $T_\textrm{c0}$ = 13.4 K. (The fit parameter $\gamma$ is related to the fit parameter $\gamma_{[31]}$ in \cite{finkelstein} by $\gamma=-1/\gamma_{[31]}$.) Inset: Variation of $T_\textrm{c}d$ with $R_{\Box}$; the line shows a fit to $T_\textrm{c} \textrm{(K)}.\,d\textrm{(nm)}=A[R_{\Box} (\Omega)]^{-B}$ as applied in Reference \citenum{ivry}.}
\end{center}
\end{figure*}

We have studied the variation of the NbN properties away from the bulk properties as we reduce the thickness dimension \cite{constantino}. Figure \ref{NCfilms}a shows the variation of sheet resistance $R_{\Box}$ with temperature for different film thicknesses in the range 10--103 nm. (Film thicknesses were measured by a stylus surface profiler.) Figure \ref{NCfilms}b shows the same data on an expanded temperature scale and normalised resistance scale, revealing that $T_\textrm{c}$ decreases as the film thickness decreases. We have focussed our study on a composition for which the thicker films have a $T_\textrm{c}$ not substantially less than the maximum obtainable $T_\textrm{c}$ for NbN, but which also displays increasing resistance as $T$ decreases from room temperature towards $T_\textrm{c}$. This shows that the films are approaching the superconductor--insulator transition. As the thickness decreases, the resistivity (not shown) calculated using the measured film thickness increases, suggesting that the thinner films are closer to the superconductor--insulator transition.
  
The progressive suppression in $T_\textrm{c}$ in thinner films is in line with observations of other materials systems \cite{crauste, ivry}. For CQPS applications, it is important to control both $T_\textrm{c}$ and $R_{\Box}$, and therefore we consider the variation of these further. In Figure \ref{NCfilms}c, we show the variation of $T_\textrm{c}$ with $R_{\Box}$, along with a fit to a model \cite{finkelstein} of $T_\textrm{c}$ suppression approaching the superconductor--insulator transition within the so-called ``fermionic scenario''. A good fit is obtained, and a value of the elastic scattering time of 3.9 fs may be extracted from the fit. This is within an order of magnitude of other values reported in the literature, although larger than other reports. For thin films, Ivry \textit{et al.} \cite{ivry} conducted a meta-analysis using results from several groups on thin films close to the superconductor--insulator transition in different materials including NbN and have empirically parametrised the behaviour in terms of parameters $A$ and $B$ through the relation $T_\textrm{c} \textrm{(K)}.\,d \textrm{(nm)}=A[R_{\Box} (\Omega)]^{-B}$ where, for unremarkable constant resistivity and $T_\textrm{c}$, a coefficient $B=1$ would be expected. The inset to Figure \ref{NCfilms}c shows that our data also fit well to this relation, with coefficients $A=(1.2\substack{+1.3 \\ -0.6}) \times 10^5$ and $B=1.04\pm 0.12$. (Note that the values obtained are not very sensitive to the precise definition used for $T_\textrm{c}$.) It is interesting that the value of $A$ we obtain is almost an order of magnitude larger than expected according to the apparent universal exponential dependence of $A$ on $B$, which Ivry \textit{et al.} \cite{ivry} arrived at empirically through the meta-analysis, compared with scatter of around half an order of magnitude or less in $A$ for a given value of $B$ for a typical sample they analysed. Our result may suggest that departures from the ``universal'' dependence are more common than indicated by that meta-analysis. 

The results in Figure \ref{NCfilms} demonstrate that sheet resistances $\gtrsim$1 k$\Omega$ may be obtained while maintaining $T_\textrm{c}>6$ K. Such values are suitable for nanowires intended to function as CQPS elements. For operating temperatures well below 1 K, optimal materials to use for CQPS applications may be films thinner than 10 nm such that the $T_\textrm{c}$ is lower, but the sheet resistance is even higher. However, due to concerns about the uniformity of thinner films and electrical continuity in those films, we chose to study nanowires fabricated from films with thicknesses in the range 10--20 nm.

\subsection{From Thin Films to Narrow Nanowires: Nanowire Fabrication}
In this section, we describe how we have obtained narrow nanowires from films of niobium nitride and in the following sections present first the variation in the properties of the niobium nitride as they are formed into nanowires of decreasing width dimension and then the characteristics of nanowires with widths down to $\sim$20 nm.
\unskip
\begin{figure*}[!ht]
\begin{center}
\includegraphics[width=12 cm]{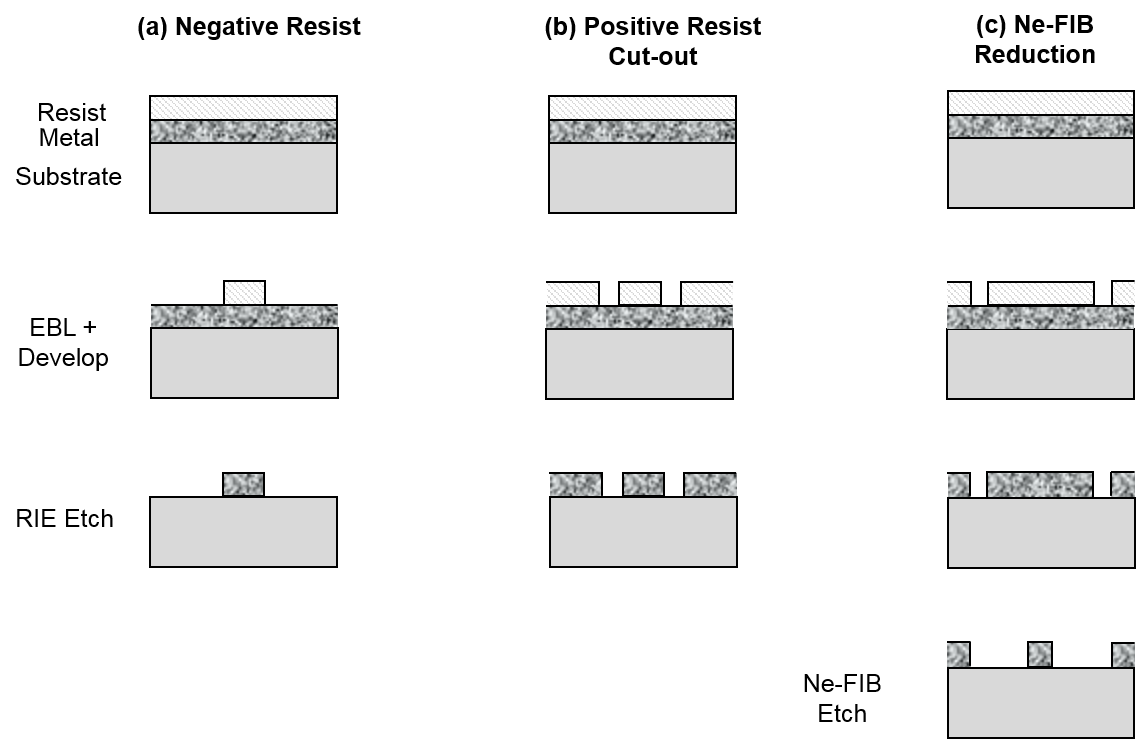}
\caption{\label{fabtech} Schematic side view cross-sections through nanowire showing the fabrication technologies employed. (\textbf{a}) Fabrication by negative resist, electron-beam lithography (EBL) and reactive ion etching (RIE). (\textbf{b}) Nanowire cut-out using positive resist, EBL and RIE. (\textbf{c}) Nanowire definition by neon focussed ion-beam milling (Ne-FIB).}
\end{center}
\end{figure*} 

\begin{figure*}[!ht]
\begin{center}
\includegraphics[width=5 cm]{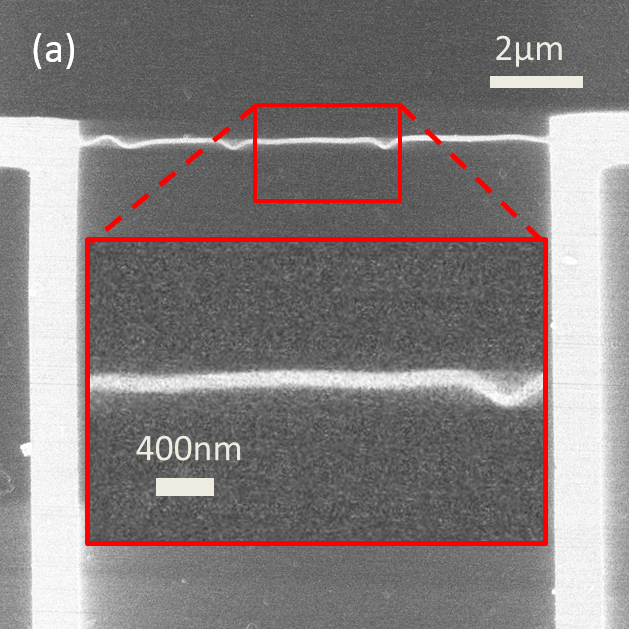}
\includegraphics[width=5 cm]{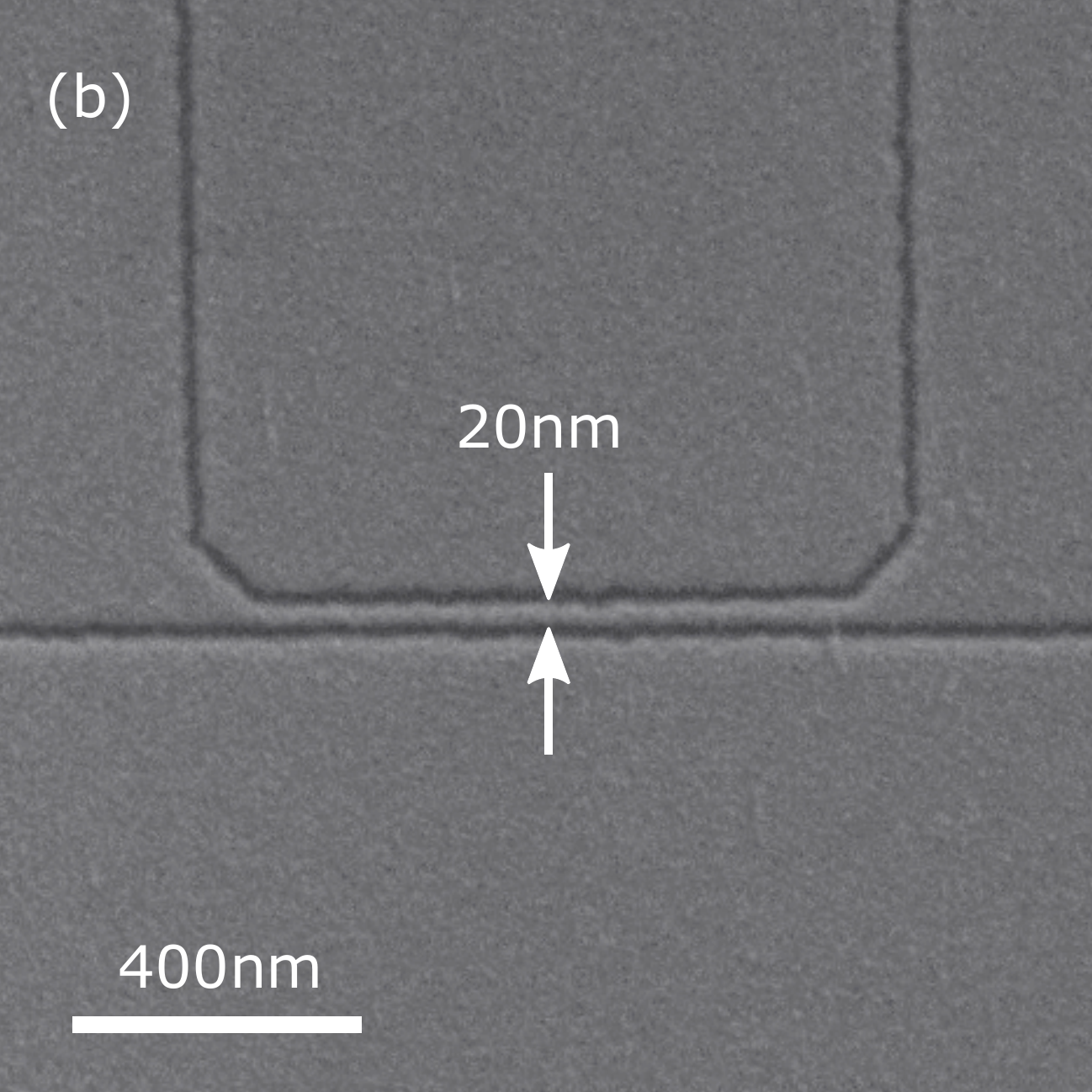}
\includegraphics[width=5 cm]{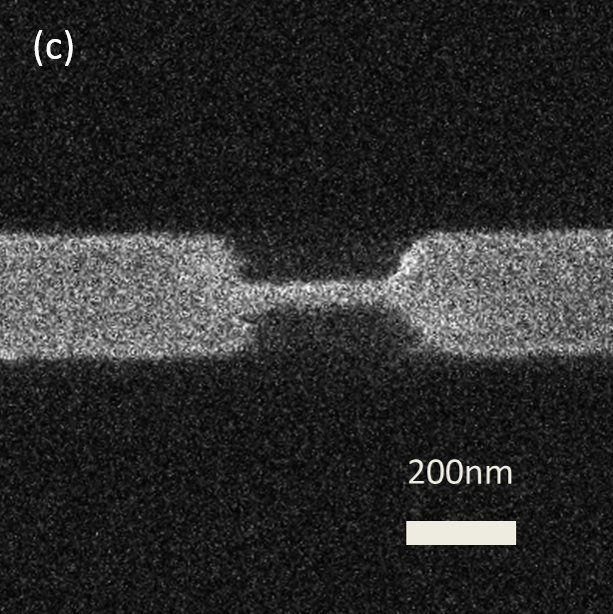}
\caption{\label{HeFIB}Plan-view micrographs of nanowires we have fabricated via three different processes. (\textbf{a})~He-FIB image of a nanowire fabricated using a hydrogen silsesquioxane (HSQ) negative-resist mask. Kinks in the nanowire shape may arise from strain relief of the nanowire resist mask during processing \cite{fentoneucas}. Inset: An enlarged view of the part of the image indicated by the upper red box. (\textbf{b})~Scanning electron micrograph of a nanowire fabricated by cut-out using a polymethyl methacrylate (PMMA) mask. (\textbf{c}) He-FIB image of a nanowire defined by Ne-FIB. In all images, light contrast shows the niobium nitride, and dark contrast shows the substrate. Note that not all the images show the narrowest nanowire obtained using this strategy.}
\end{center}
\end{figure*}

There are several options for defining very narrow nanowires. We have used three different techniques to define narrow nanowires, based either on electron beam lithography or on neon focussed ion-beam milling (Ne-FIB). These are shown schematically in Figure \ref{fabtech} and described in detail in the Materials and Methods section, but summarised briefly here, and examples of nanowires obtained using all three strategies are shown in Figure \ref{HeFIB}.
A first technique we have used to define very narrow nanowires via an electron-beam lithography (EBL) stage is to use a negative-tone e-beam resist to define a mask. Single-pixel lines are exposed in the resist, and this pattern is transferred to the film by reactive ion etching (RIE). 
A second technique we have used for defining nanowires utilises a positive-tone resist rather than a negative resist. A ``cut-out'' strategy \cite{constantino} is used, in which we use EBL to expose the resist to define the edges of the nanowires to be removed, then transfer the pattern to the film by RIE. 
A third fabrication strategy we have employed does not rely on either an e-beam resist mask or RIE when defining the nanowire. We use focussed neon ion-beam milling, which has resolution as good as 5 nm, in order to remove material from the film to define the nanowire.

\subsection{Spectrum of Nanowire Properties}
\begin{figure*}[!ht]
\begin{center}
\includegraphics[width=5.1 cm]{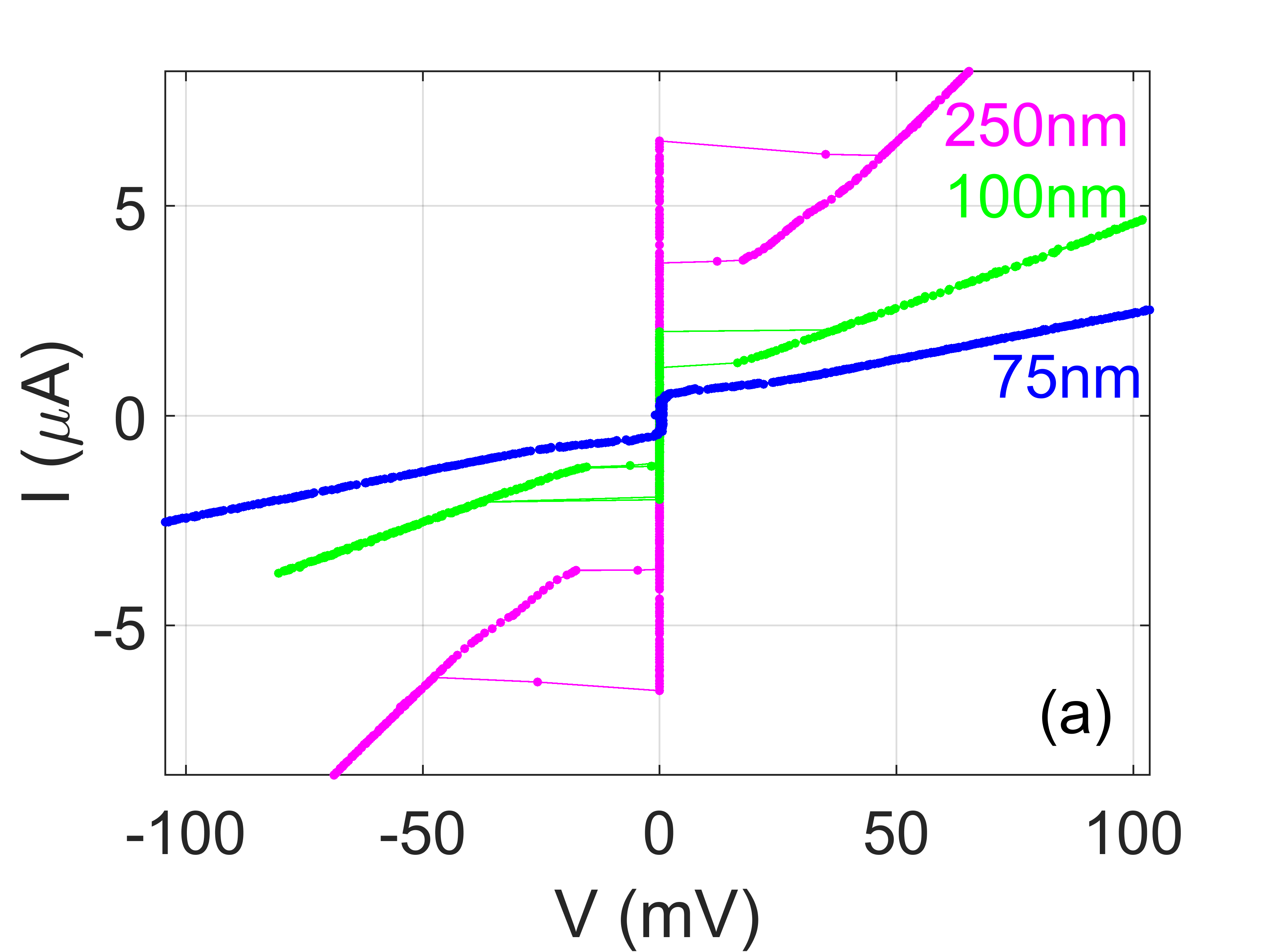}
\includegraphics[width=5.1 cm]{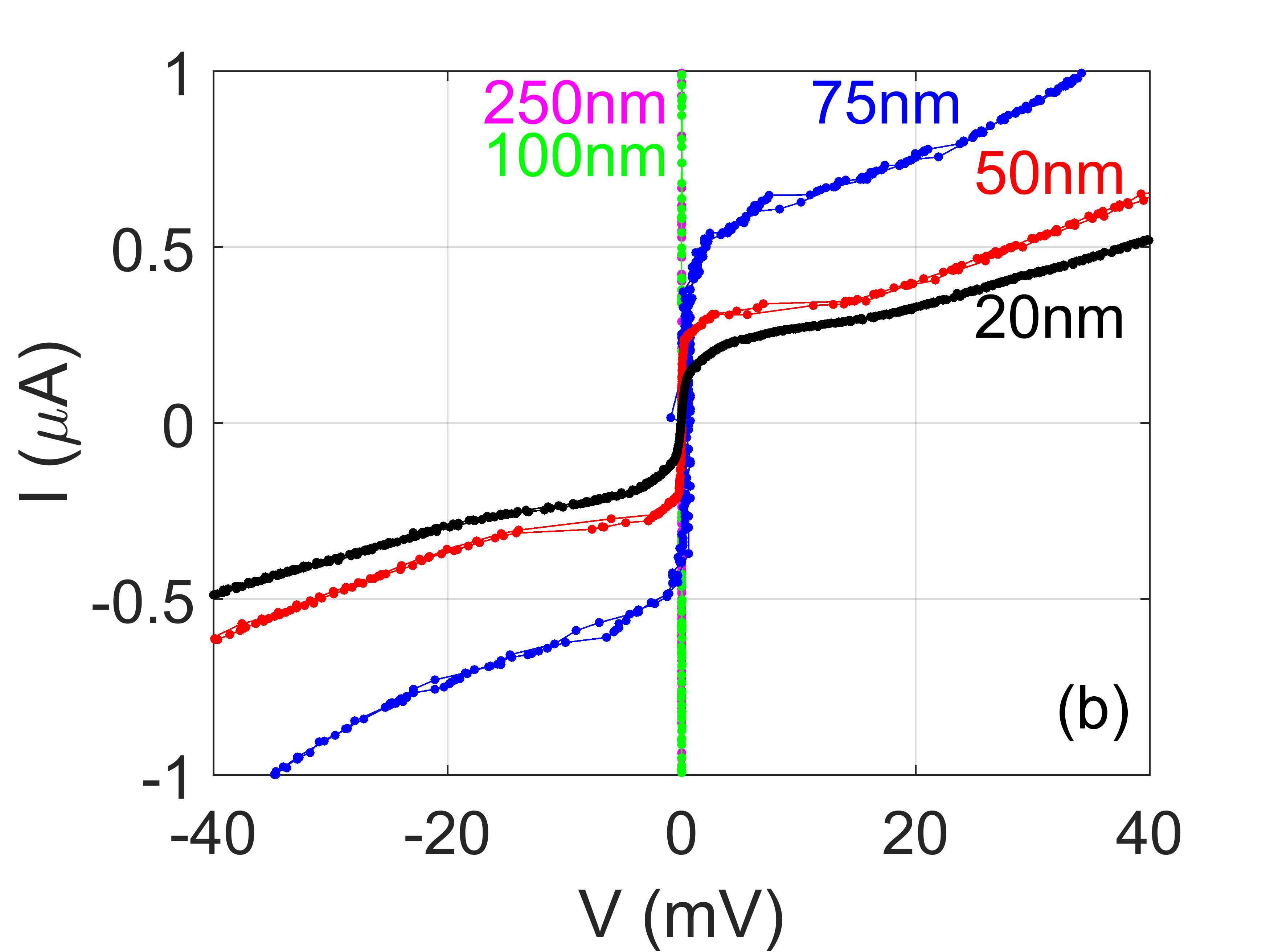}
\includegraphics[width=5.1 cm]{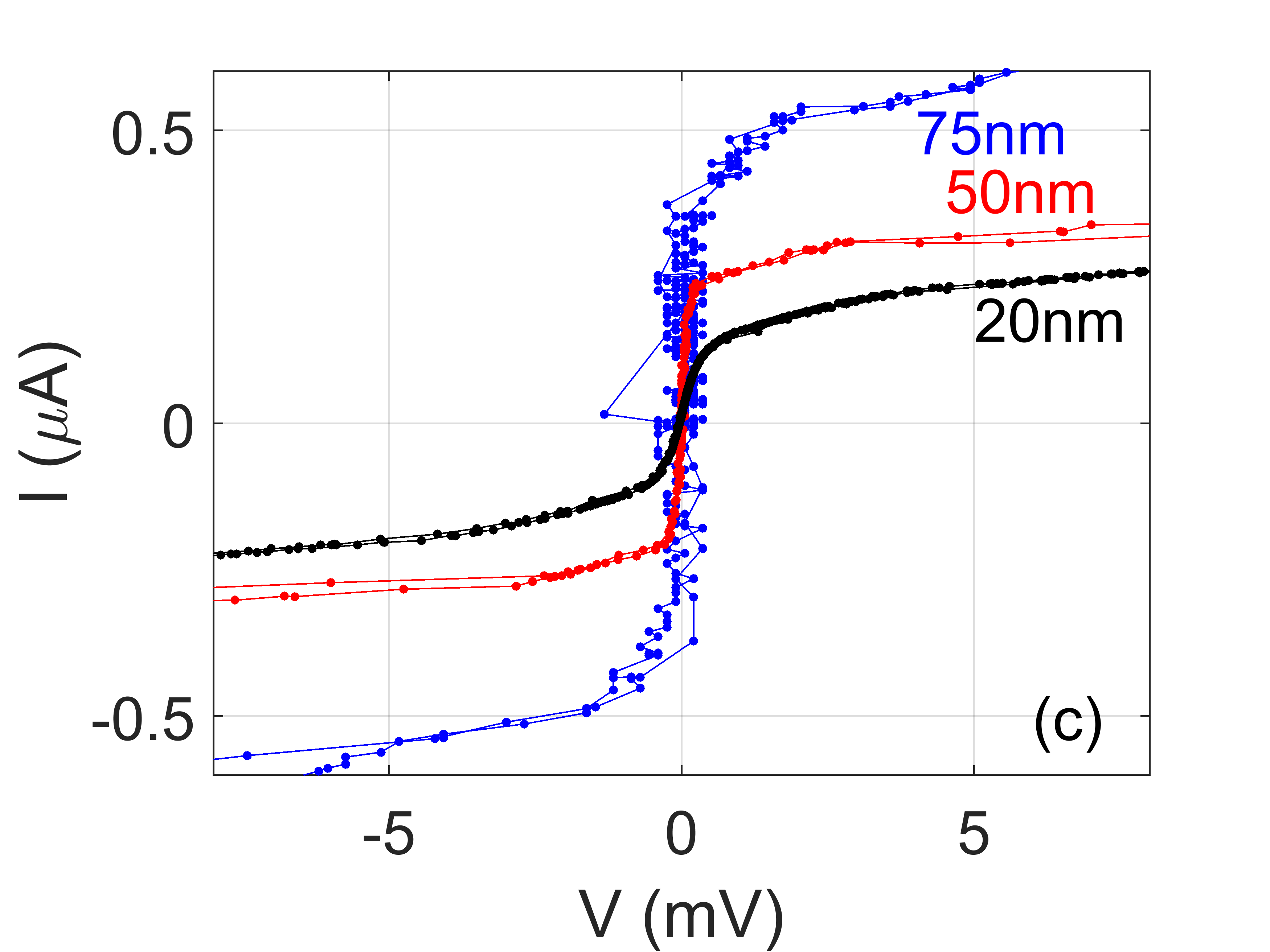}
\caption{\label{NCnwIVs}$I(V)$ measurements at 4.2 K for five nanowires with widths in the range 20--250 nm in a film of a thickness of 10 nm. (\textbf{a}) $I(V)$ at 4.2 K for nanowires of widths of 250 nm, 100 nm and 75 nm. Lines~join consecutive points. (\textbf{b}) Same data zoomed-in at low bias and also showing data for nanowires of widths of 50 nm and 25 nm. (\textbf{c}) Data for the three narrowest nanowires on a further expanded scale. Notice that the data for the 50-nm-wide and 75-nm-wide nanowires contain small jumps for $V<20$ mV.}
\end{center}
\end{figure*}

\begin{figure*}[!hbt]
\begin{center}
\includegraphics[width=4.5 cm]{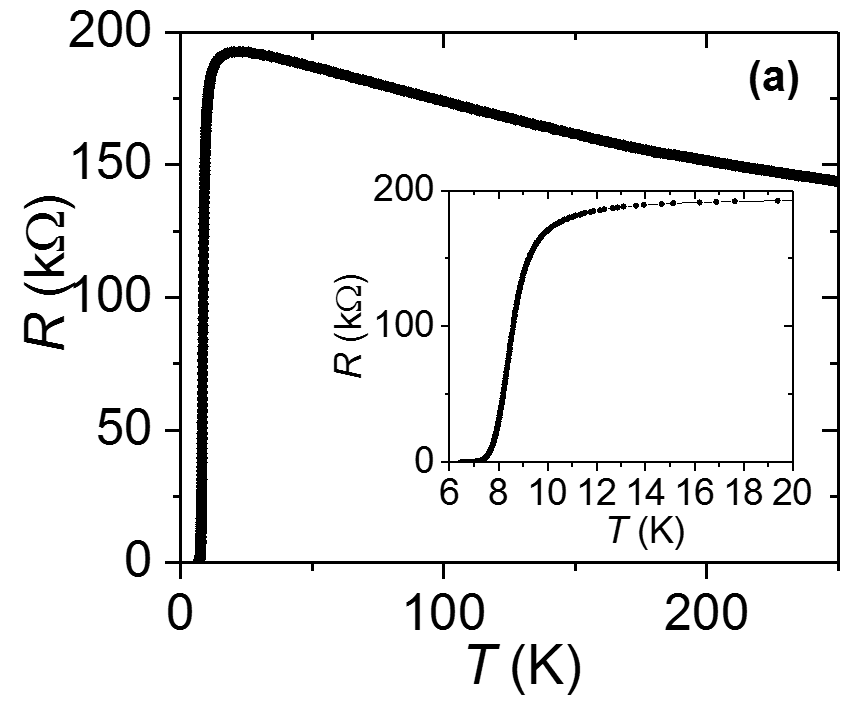}
\includegraphics[width=5.6 cm]{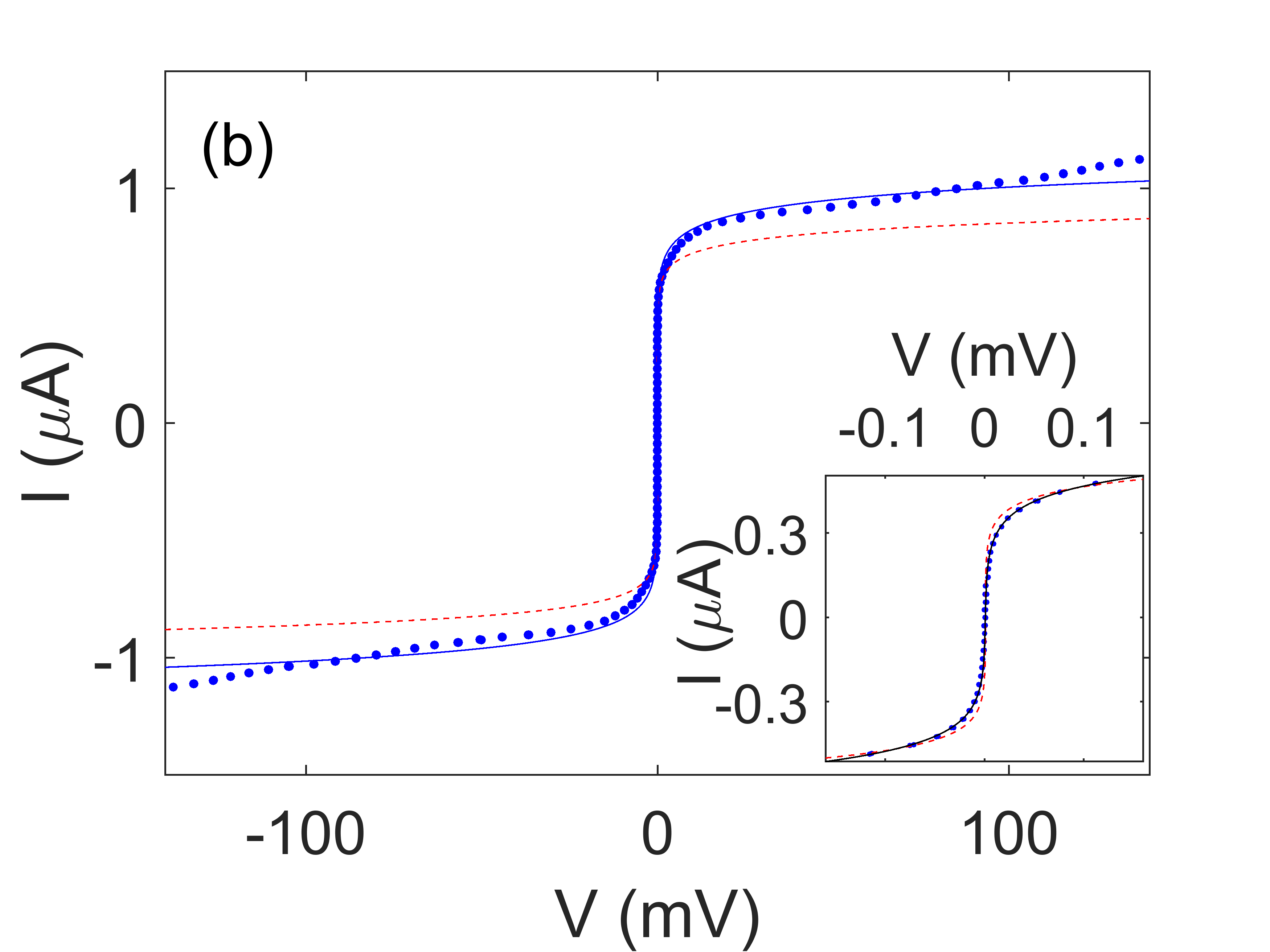}
\caption{\label{NbN81cell2}Measurements on sample NbN81/2. (\textbf{a}) $R(T)$. The inset shows the same data on an expanded scale. (\textbf{b}) $I(V)$ at 330 mK. The lines show fits to the form $V=V_i\sinh{(I/I_i)}$: the dashed line shows the best fit to thermally activated phase-slips (TAPS), with $V_\textrm{TAPS}=45.7$ mV, and the dashed line shows the best fit as a model of incoherent quantum phase-slips (IQPS), with $V_\textrm{IQPS}=467$ mV and $I_\textrm{IQPS}=77.9$ mA. The inset shows the same data on an expanded scale. }
\end{center}
\end{figure*}

\begin{figure}[!bht]
\begin{center}
\includegraphics[width=7 cm]{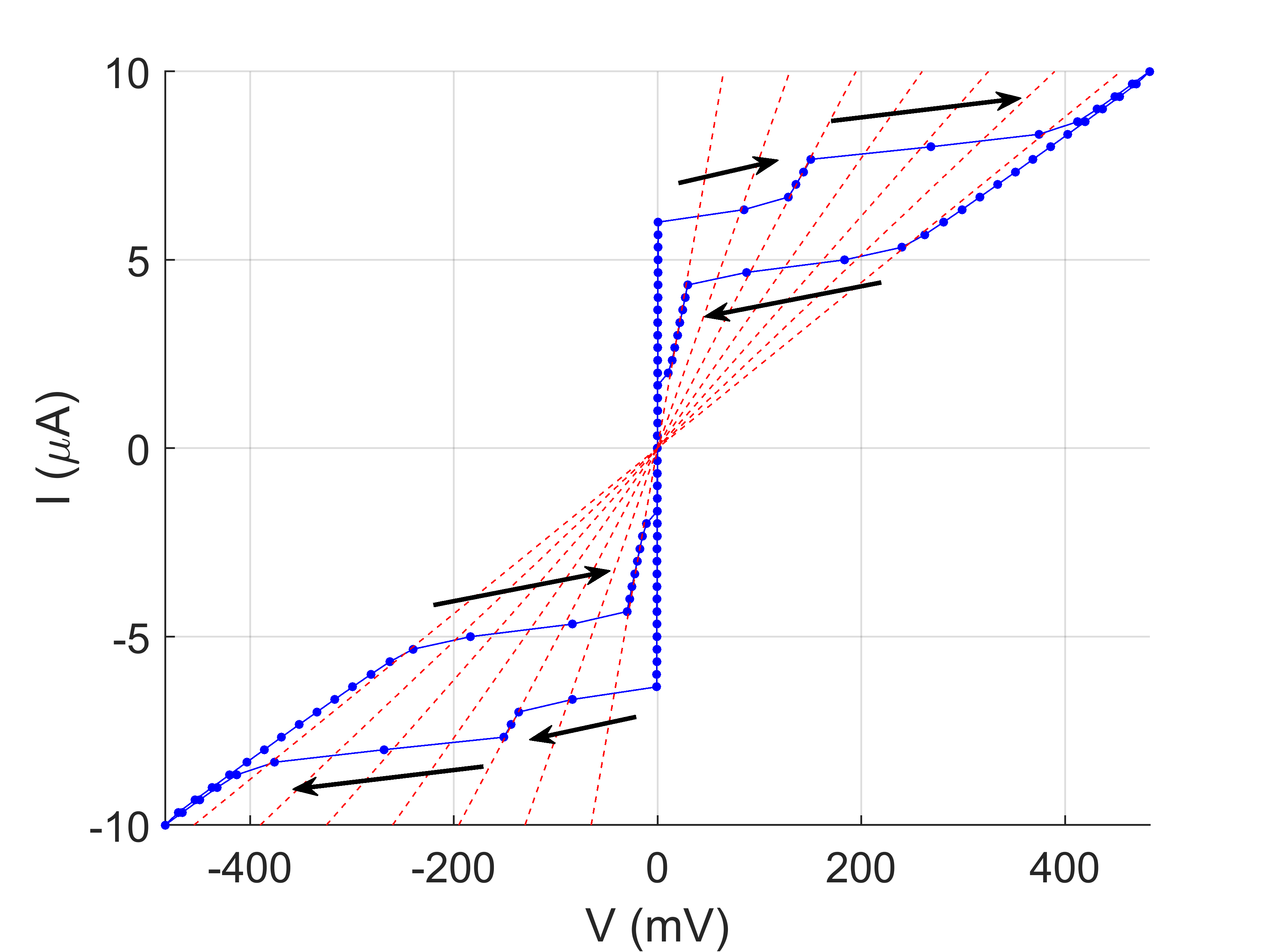}
\caption{\label{PSC}$I(V)$ for nanowire NbN65/1 at 320 mK, which has a width of 40 nm and a length of 220 nm and was fabricated using Ne-FIB. The $I(V)$ shows possible phase-slip-centre behaviour. Solid lines join consecutive points; arrows show the direction of current sweep; and dashed red lines show resistance multiples of 6.5 k$\Omega$.}
\end{center}
\end{figure}

We will now present DC electrical characterisation of NbN nanowires obtained by all three techniques. $I(V)$ measurements at cryogenic temperatures on several NbN nanowires are shown in Figures \ref{NCnwIVs}--\ref{NbN80cell1}, and these show a range of different behaviours. Relevant parameters of these nanowires are listed in Table \ref{table}.

Figure \ref{NCnwIVs} shows $I(V)$ measurements at 4.2 K on nanowires with widths in the range 20--250 nm, fabricated on a film of a thickness of 10 nm using the cut-out technique. The $T_\textrm{c}$ of all these nanowires is around 7 K. The shape of the $I(V)$ changes qualitatively depending on the width of the nanowire. As Figure \ref{NCnwIVs}a shows, for nanowires with widths of 250 nm or 100 nm, the low-bias resistance is zero, and there is a switch at a certain critical current to a state with an approximately constant resistance. When the current is subsequently decreased, the nanowire returns to a zero-resistance state at a lower current. A similar hysteretic behaviour has often been observed in superconducting nanowires and is generally the result of electronic heating due to dissipation in the nanowire, raising its temperature and so suppressing its critical current.

Figure \ref{NCnwIVs}b shows that the nanowires with widths of 75 nm and 50 nm, in contrast, display barely any hysteresis and also a gradual increase in resistance as current increases rather than a single jump to a large resistance. On close inspection of this range, several jumps in voltage of $\approx$1--7 mV may be observed in the voltage range 3--25 mV. The cause of these branches is not completely clear; the branches resemble both the $I(V)$ that would be obtained in an array of Josephson junctions with some distribution in critical currents and somewhat resemble the phase-slip centre behaviour that is sometimes observed in nanowires (phase-slip centre behaviour will be discussed in more detail below in relation to Figure \ref{PSC}). Let us compare the size of the jumps to what would be expected for Josephson junctions forming between grains in a superconducting film: For a BCS superconductor $I_\textrm{c}R_\textrm{N}\sim \pi\Delta/(2e)$ at low temperature with the superconducting energy gap $\Delta =1.76k_\textrm{B}T_\textrm{c}$, which for $T_\textrm{c}=8$ K would give $I_\textrm{c}R_\textrm{N}$=1.9 mV (or perhaps a factor of $\sim$2 larger for a non-BCS superconductor as a result of a larger $\Delta$) \cite{tinkham} and typically high-quality Josephson junctions display such values of $I_\textrm{c}R_\textrm{N}$, while less carefully prepared Josephson junctions display smaller values. Given the possibility that some observed jumps may involve more than one Josephson junction becoming resistive at once, the~$\approx$1--7~mV observed here therefore appears consistent with grain-boundary Josephson junctions having formed in these particular nanowires and being responsible for the observed jumps.

The narrowest nanowire has a non-hysteretic $I(V)$ and does not display any jumps. The low-bias resistance is still non-zero, increasing as the current increases. This qualitative behaviour is one of the typical characteristic behaviours we observe in other NbN nanowires (see Figure \ref{NbN81cell2}; data for the other nanowires is not shown). The appearance of resistance is reminiscent of much wider superconducting wires, in which resistance appears due to vortex flow. However, these nanowires are narrower than the Josephson penetration depth, and therefore vortex flow cannot be responsible.
Extrapolating from discussion for the 50-nm and 75-nm nanowires, one conceivable explanation is that these two nanowires are acting as an array of Josephson junctions with a particular smoothly varying distribution of critical currents. If the Josephson junctions were overdamped, this could explain the absence of discernible jumps.
The observed behaviour is also qualitatively as expected for a nanowire undergoing TAPS or incoherent quantum phase-slips. Both TAPS and incoherent QPS (IQPS) lead to an increasing voltage at higher current as the current through the nanowire reduces the energy barrier for a phase-slip, and this can be expressed as $V=V_i\sinh{(I/I_i)}$, where $I_i$ is related to the energy barrier and $i$ denotes either TAPS or IQPS \cite{altomare}. For TAPS, $I_\textrm{TAPS}=4ek_\textrm{B}T/h$, whereas, for IQPS, $I_\textrm{IQPS}$ is temperature-independent. Quantitative fits based on a single value for $I_\textrm{IQPS}$ do not generate a satisfactory fit unless the data range fitted is artificially restricted substantially. However, if instead there was a particular distribution of energy barriers for phase-slips within the wire, leading to an incoherent QPS voltage made up of a sum of the terms with a range of $I_\textrm{IQPS}$ values, this could also quantitatively explain the $I(V)$.

Figure \ref{NbN81cell2} shows the $R(T)$ of a different NbN nanowire, fabricated by the negative-resist technique with HSQ and having a nominal thickness of 18 nm and a width of around 60 nm. The resistance has a maximum at 22.5 K corresponding to a sheet resistance of $\approx$1 k$\Omega$ per square. The ratio of the resistance at 200 K to the maximum low-temperature resistance is $\approx$1.25, approximately the same as for the films shown in Figure \ref{NCfilms}a, suggesting that the reduction in nanowire width does not significantly further increase the approach to the superconductor--insulator transition seen on decreasing film thickness in Figure \ref{NCfilms}b. The superconducting transition occurs at around 9 K, with a width of 2.5 K. Minimal~residual resistance below $T_\textrm{c}$ is seen in Figure \ref{NbN81cell2}a.
The low-temperature $I(V)$ for this nanowire is qualitatively similar to the data for the narrowest nanowire in Figure \ref{NCnwIVs}. Figure \ref{NbN81cell2}c shows fits to the models of TAPS and incoherent QPS described earlier, with the range of values for the fit restricted to $|V|<160$ $\upmu$V. The fit to the TAPS model is poor over the entire range of the data, but ---unlike for the narrowest nanowire in Figure \ref{NCnwIVs}--- the incoherent QPS model gives an excellent fit at low bias. In~addition, the fit also gives reasonable agreement with the data for $|V|>160$ $\upmu$V, except at the highest bias where it is quite possible that other physics is affecting the behaviour. This suggests that, in this nanowire, incoherent QPS may well be responsible for the voltage developed at low currents.

Figure~\ref{PSC} shows an $I(V)$ characteristic of another nanowire, NbN65/1, measured at 330 mK. This~nanowire was generated by neon FIB milling and has a width of 40 nm. As the current is swept first from 0 to 10 $\upmu$A, then back down to $-10$ $\upmu$A before returning to zero, resistive branches on which $\textrm{d}V/\textrm{d}I$ is approximately constant are seen. The resistive branches are at integer multiples of a fraction of the nanowire's normal-state resistance, and this matches the characteristic behaviour of phase-slip centres. Phase-slip centres ---distinct from the isolated phase-slips previously described--- are periodic order parameter oscillations at one or more locations along the nanowire and lead to dissipation, both from the oscillating region and from adjoining regions in which a nonequilibrium population of quasiparticles decays over a length scale much longer than the superconducting coherence length \cite{tinkham}. In this measurement, there are clear branches at certain multiples $n$ of 6.5 k$\Omega$, suggesting that a single phase-slip centre contributes a resistance of 6.5 k$\Omega$, with $n$ phase-slip centres being found along the nanowire. In previous reports, phase-slip centres have often been observed to form at temperatures approaching $T_\textrm{c}$; however, this measurement was carried out at 320 mK, well below $T_\textrm{c}\sim 4$ K. (The composition of this sample was different from the others we report here. It was sputtered from a target with a different composition and contained a significant proportion of tantalum. This is likely to be the reason for the lower $T_\textrm{c}$.) This nanowire was one of the first we prepared using Ne-FIB for DC transport measurements, and only this one of the first eight comparable nanowires prepared similarly showed this phase-slip-centre behaviour. Further investigations of DC-transport behaviour in Ne-FIB-fabricated nanowires are in progress.

\subsection{Coherent Quantum Phase-Slip Behaviours in NbN Nanowires}
\begin{figure*}[!htb]
\begin{center}
\includegraphics[width=5 cm]{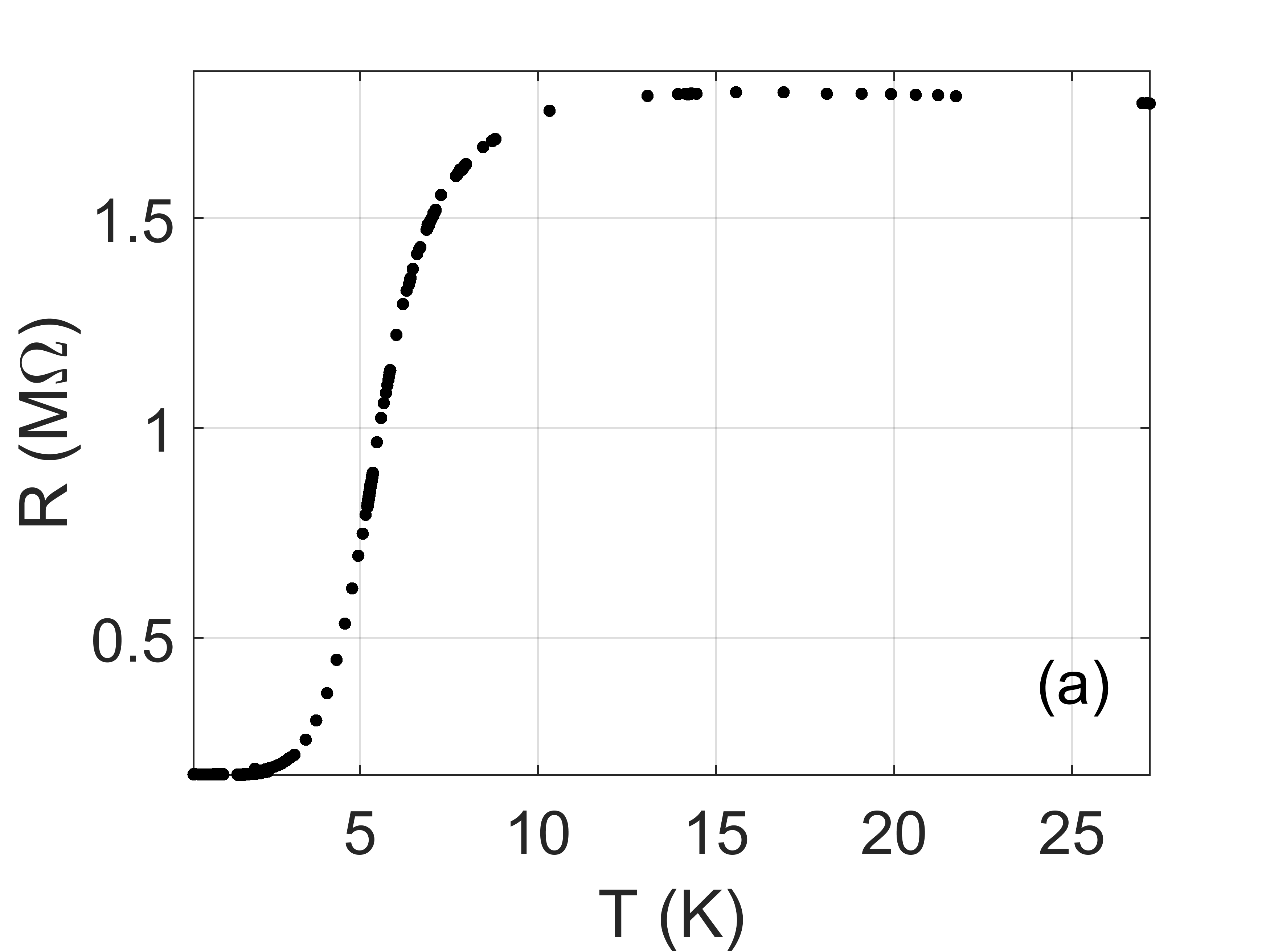}
\includegraphics[width=5 cm]{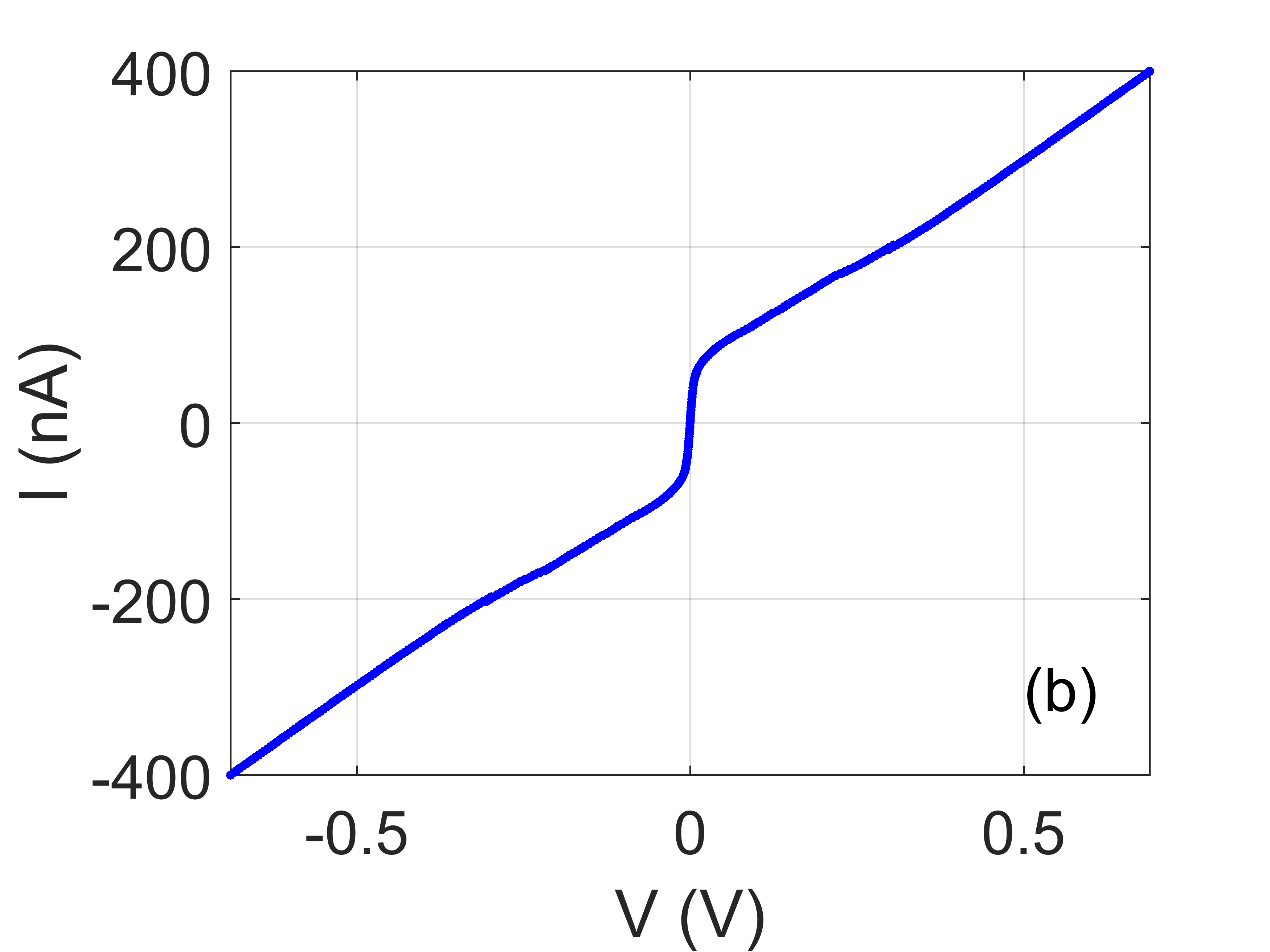}
\includegraphics[width=5 cm]{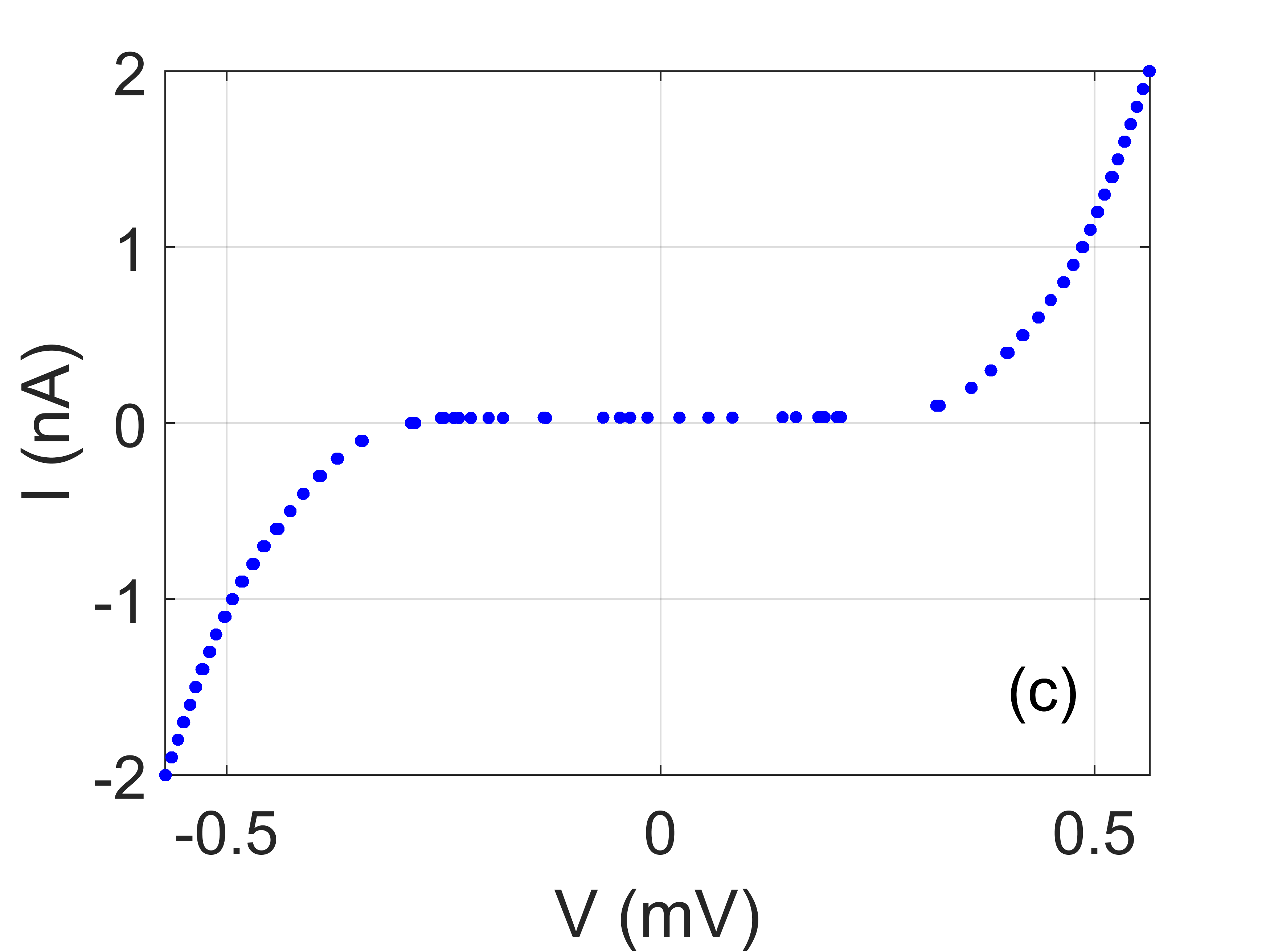}
\caption{\label{NbN25/A1}Measurements of sample NbN25/A1. (\textbf{a}) $R(T)$, found by fitting to $I(V)$ taken with \mbox{$|I|\leq 10$ nA}. (\textbf{b}) $I(V)$ at 330 mK. (\textbf{c}) Low-bias sweep at 330 mK, showing a critical voltage feature.}
\end{center}
\end{figure*} 
\begin{figure*}[!ht]
\begin{center}
\includegraphics[width=5.1 cm]{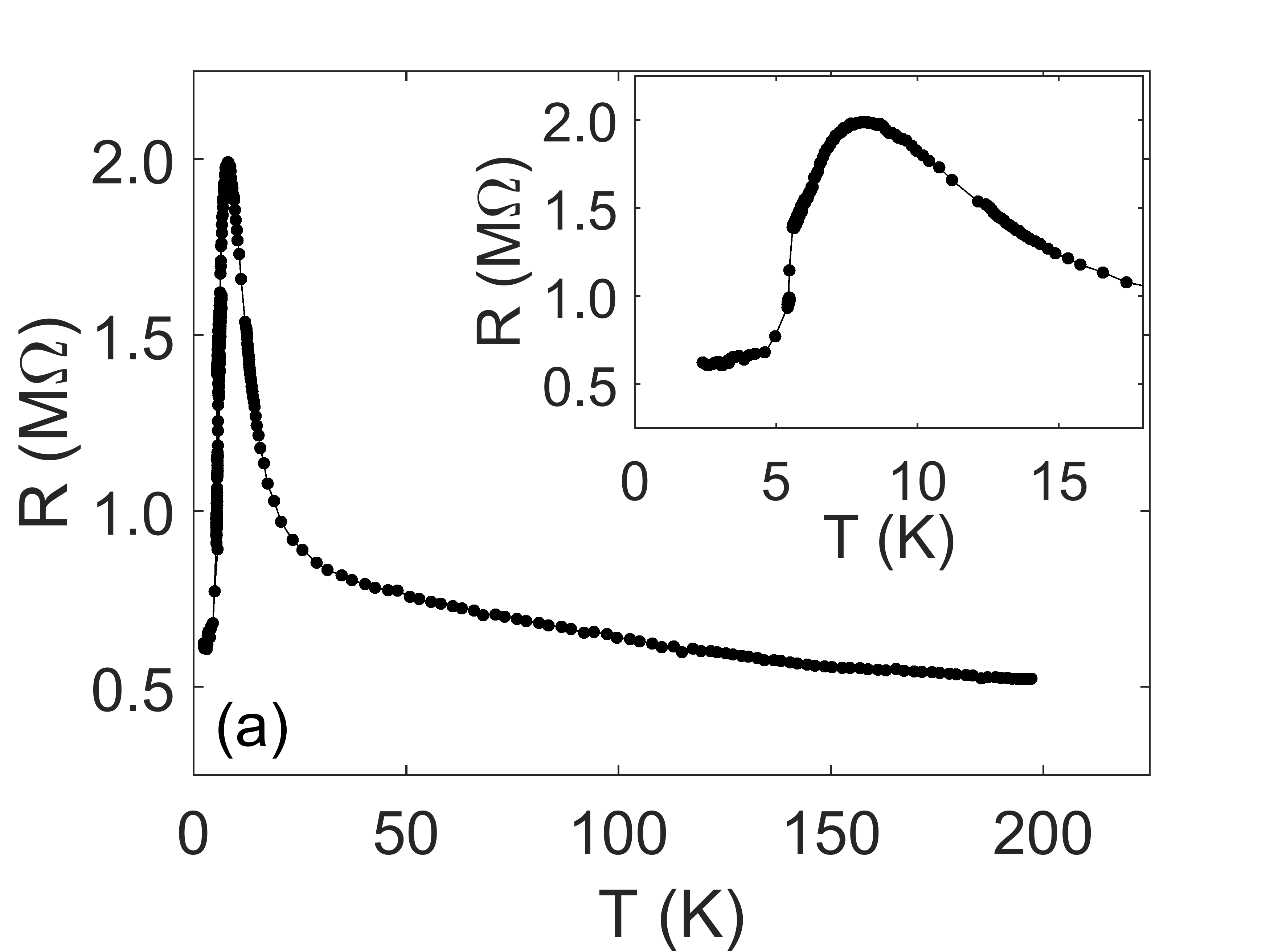} 
\includegraphics[width=5.1 cm]{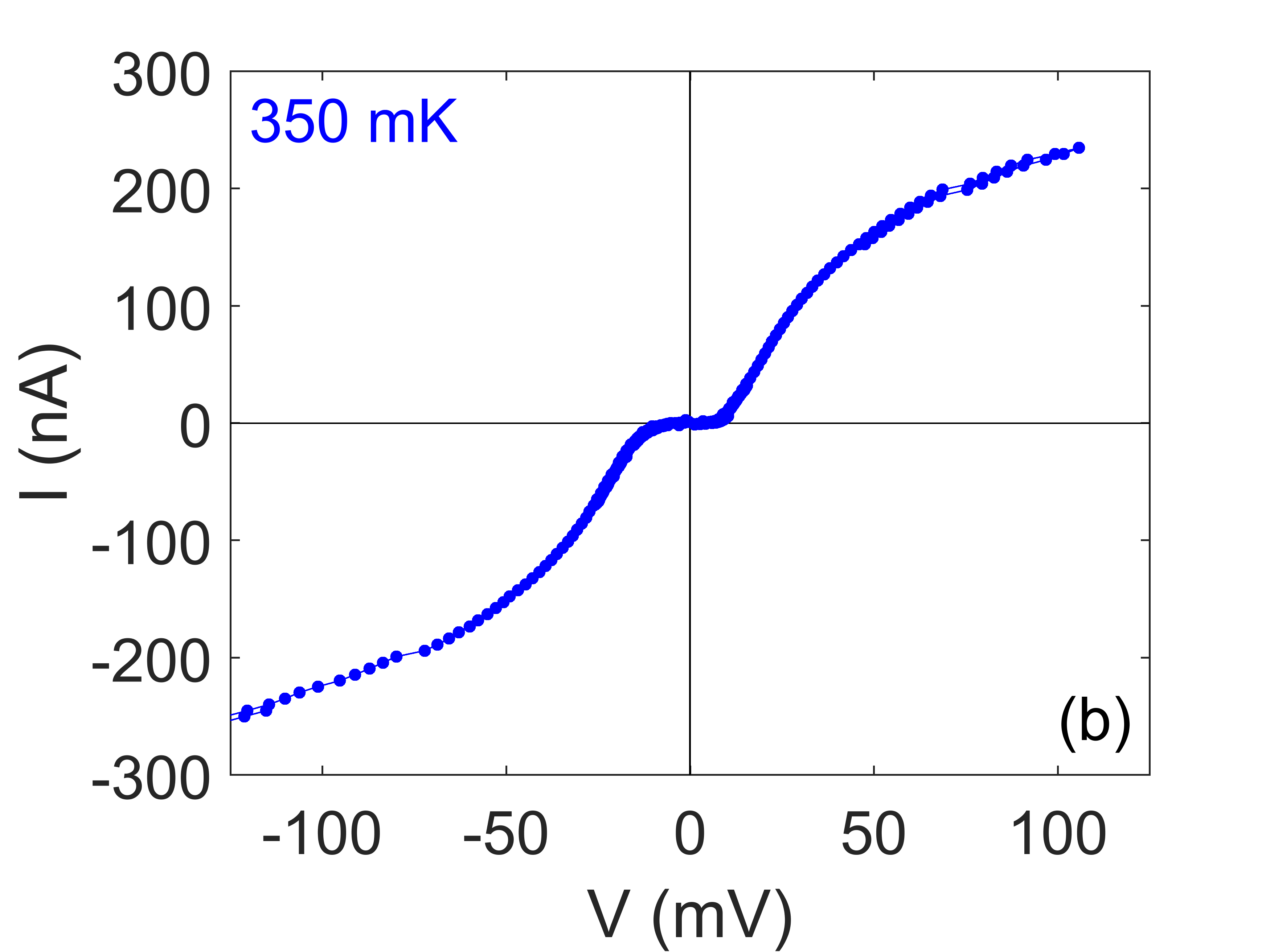} 
\includegraphics[width=5.1 cm]{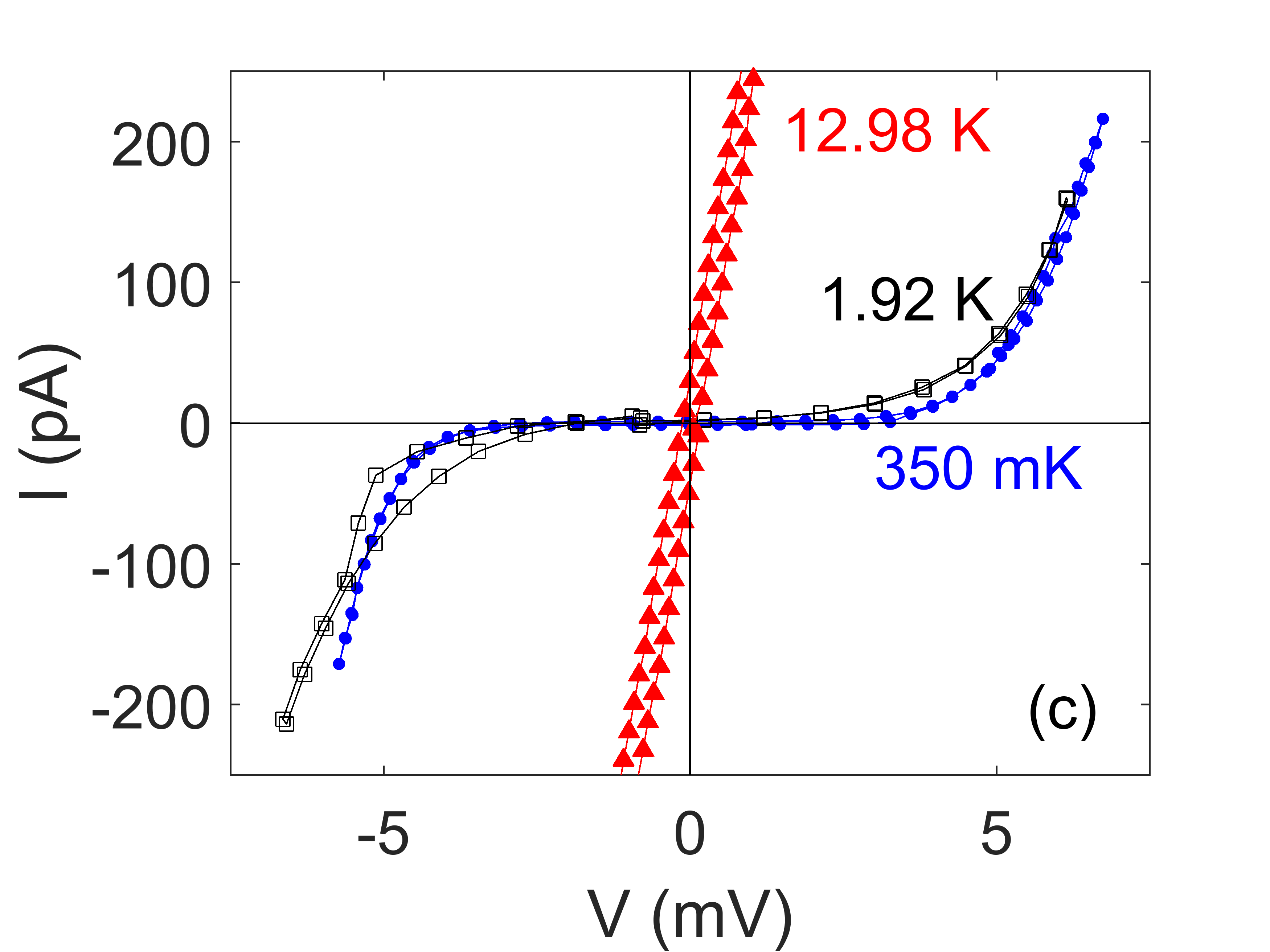} 
\caption{Measurements on sample NbN80/1. (\textbf{a}) $R(T)$. Inset: the same data on an expanded scale showing the low-temperature region. (\textbf{b}) High-bias $I(V)$ at 330 mK. (\textbf{c}) $I(V)$ at three temperatures above and below $T_\textrm{c}$, 350 mK, 1.92 K and 12.98 K. In all sub-plots, lines join consecutive data points. Voltage- and current-offsets of $-$1.25 mV, 0.5 mV and $-$0.28 mV and 5.5 pA, 6.0 pA and 5.5 pA, respectively, have been subtracted from the respective datasets in (c). The slight hysteresis observed in the measurement at 12.98 K is not a property of the sample, but rather an artefact associated with carrying out the measurement relatively rapidly (see Section \ref{meastdisc}).} \label{NbN80cell1}
\end{center}
\end{figure*} 

Figure \ref{NbN25/A1} shows $R(T)$ for a NbN nanowire with a width of $\approx 50$ nm and a length of 18 $\upmu$m, created using the negative-resist technique. Figure \ref{NbN25/A1}a shows a measurement of $R(T)$ below 25 K. The nanowire has $T_\textrm{c}\approx 6$ K and a rather broad width $\approx 4$ K. The maximum resistance, 1.8 M$\Omega$, implies a sheet resistance $\approx$3 k$\Omega$, which is rather high. There is an apparent residual resistance below $T_\textrm{c}$, and this can be better understood by examining the $I(V)$ dependence: Figure \ref{NbN25/A1}b shows $I(V)$ at 330 mK, and this has a similar shape to Figure~\ref{NbN81cell2}; however, closer inspection of the low-bias region (Figure \ref{NbN25/A1}c) reveals an additional region at low bias, in which no current is measured below a critical voltage \mbox{$V_\textrm{c}\approx 300$ $\upmu$V.} This behaviour is suggestive of the presence of coherent quantum phase-slips, and $V_\textrm{c}$ is similar to previous reports \cite{websterfenton,hongisto}. This value of $V_\textrm{c}$ implies that $E_\textrm{S}\approx 3.3k_\textrm{B}T$ at 330 mK, suggesting that the characteristic voltage of this nanowire is just large enough not to be substantially thermally rounded at this temperature.
 
Figure \ref{NbN80cell1} shows the results of characterisation of another nanowire, with a length of 9 $\upmu$m, an approximate width of 60 nm and a nominal thickness of 18 nm, created by the negative-resist technique. This nanowire was embedded in an environment with NbN series inductors with length 400 $\upmu$m and width 200 nm and series resistance 148 k$\Omega$. Figure \ref{NbN80cell1}a shows that there is a steeper increase in the resistance at a low temperature, indicating that this composition is much more resistive than those shown in Figure~\ref{NCfilms}, and the maximum sheet resistance based on the physical dimensions is 10 k$\Omega$. The sample is superconducting with \mbox{$T_\textrm{c}\approx$ 5.5 K}, with a broad superconducting transition of a width of $\approx$2--3 K. Figure \ref{NbN80cell1}c shows that a clear critical voltage feature develops below $T_\textrm{c}$, with $V_\textrm{c}\approx$ 5 mV at 350 mK. This is an order of magnitude larger than the earlier reports of a $V_\textrm{c}$ feature in NbSi nanowires \cite{websterfenton,hongisto}. Figure \ref{NbN80cell1}c also shows that the critical voltage feature is a little rounded at 1.92 K and is completely absent above $T_\textrm{c}$.

Tangents to the high-bias part of the $I(V)$ characteristics in both Figures \ref{NbN25/A1}b and \ref{NbN80cell1}b intercept the current axis at a positive current. This is an important feature of the observed behaviour \cite{websterfenton}. The behaviour is characteristic of superconductivity; by contrast, for a current blockade arising from single-electron effects, a tangent to the higher bias part of the $I(V)$ characteristic would be expected to intercept the current axis at a negative current. The positive current intercept and the development of the $V_\textrm{c}$ feature over the same temperature range as superconductivity (Figure \ref{NbN80cell1}c) indicate that the $V_\textrm{c}$ feature arises out of superconductivity rather than being independent of it. This observation is as expected if the behaviour indeed arises from coherent quantum phase-slips and would not be expected for a single-electron Coulomb blockade in a tunnel junction.

\section{Discussion}
\begin{table*}[!hbt]
\caption{Nanowire sample properties. Fab method = fabrication method, Cut-out = cut-out using PMMA resist and EBL, HSQ = fabrication by EBL using hydrogen silsesquioxane (HSQ) resist, Ne-FIB = nanowires defined by neon FIB, sc = standard superconducting behaviour, PSCs = phase-slip centres, IQPS = successful fit of $I(V)$ to a model based on incoherent quantum phase-slips (see the main text). $R_{\Box}$ is the sheet resistance above $T_\textrm{c}$. Indicated $T_\textrm{c}$ values are defined as $R(T_\textrm{c})=0.5R_\textrm{max}$, where $R_\textrm{max}$ is the maximum resistance and $\Delta T_\textrm{c}=T(R=0.9R_\textrm{max})-T(R=0.1R_\textrm{max})$. `Circuit elements' indicate thin-film components included in series in the circuit; nanowires NbN80/1 and NbN81/2 have series resistance of 148 k$\Omega$ and series inductance with a length of 400 $\upmu$m and a width of 200 nm. Nanowires 100414 and the films for which $R(T)$ is shown in Figure \ref{NCfilms} were deposited on silicon substrates and were not measured below 4.2 K; all other samples were deposited on sapphire substrates.} \label{table}
\begin{center}
\begin{tabular}{ccccccccc}
\hline
\textbf{Sample}	& \textbf{Fab Method} & \textbf{Circuit Elements} & \textbf{\boldmath{$l$} (\boldmath{$\mu$}m)} & \textbf{\boldmath{$w$} (nm)} & \textbf{\boldmath{$R_{\Box}$} (k\boldmath{$\Omega$})} & \textbf{\boldmath{$T_\textrm{c}$ }(K)}	& \textbf{\boldmath{$\Delta T_\textrm{c}$} (K)} & \textbf{Behaviour Summary}\\
\hline
100414/D2		& Cut-out		& none & 1 & 250 & $\sim$2 & $\approx$7.5 & $\approx$2 & sc\\ 
100414/D1		& Cut-out		& none & 1 & 100 & $\sim$2 & $\approx$7.5 & $\approx$2.5 & sc\\
100414/C2		& Cut-out		& none & 1 & 75 & $\sim$2 & $\approx$7.1 & $\approx$3 & sc\\
100414/C1		& Cut-out		& none & 1 & 50 & $\sim$2 & $\approx$6.7 & $\approx$3 & sc\\
100414/B2		& Cut-out		& none & 1 & 25 & $\sim$2 & $\approx$7.2 & $\approx$4 & sc\\
NbN25/A1		& HSQ		& none &	18 & 50 & $\approx$5 & 5.5 & 5 & $V_\textrm{c}\sim$300 $\upmu$V\\
NbN65/1 	& Ne-FIB & none &	0.22 & 40 & $\sim$9 & 4 & 1 & sc, PSCs\\
NbN80/1		& HSQ		&	L,R & 10.5 & $\approx$60 & $\approx$10 & 5.5 & 2--3 & $V_\textrm{c}\sim$5 mV\\
NbN81/2		& HSQ		&	L,R & 10.5 & $\approx$60 & $\approx$1 & 8.5 & 2 & sc, IQPS\\
\hline
\end{tabular}
\end{center}
\end{table*}
Table \ref{table} collects relevant parameters for several nanowires we have fabricated. It may be observed that the different types of behaviour do not show a simple correlation with variations in any of the nominal physical dimensions of the material, and we have also found that this is a typical feature of the behaviour in other nanowires we have measured (data not shown here). Such differences are usually posited to be due to variations in the cross-section of the nanowire along its length, with the CQPS properties arising from just the smallest cross-section point of the nanowire, although this is difficult to be definitive about in the absence of control and/or characterisation at the atomic scale. 

The observation of nanowire-to-nanowire variability of properties has apparent implications for technological applications. Sample-to-sample variability in properties has previously been observed in nanowire samples \cite{mooijnjp,ivry}, and variations in properties have been explained systematically according to a relationship that is not trivially determined by a single variable such as $\xi$ or $R_\textrm{N}$, but that depends on several variables. Within the analysis of Mooij \textit{et al.} \cite{mooijnjp}, whether a nanowire displays superconducting or insulating behaviour is expected to depend on both $R_\xi$ and the nanowire length $l$, with a critical resistance in the case of no additional series inductance, $R_{\xi\textrm{,crit}}(l,\xi)=b/\ln{(c(l/\xi)^2)}$, where $c=a/(17.4\alpha_\textrm{c})$ and $a$ and $b$ are the numerical constants appearing in Equation \ref{eq}. In Reference \citenum{mooijnjp}, they take $\alpha_\textrm{c}$ to be 0.3. Taking the values used in Reference \citenum{mooijnjp} for $b$ and $c$ along with $\xi$ = 5 nm, for $l$ = 1 $\upmu$m the relation gives $R_{\xi\textrm{,crit}}=0.0015R_\textrm{Q}\approx 100~\Omega$, with smaller $R_{\xi\textrm{,crit}}$ for longer nanowires. All our nanowires therefore should lie on the insulating side of this boundary, yet our nanowires do not show behaviour as insulating ({i.e.}, as large $V_\textrm{c}$) as expected for their length on the basis of this calculation. Possible explanations for this include that the values used for $b$ and $c$ should be modified for our material system or that the length within our nanowires in which CQPS is occurring is much smaller than their physical length.

It is also possible that the relevant active cross-sectional dimensions of the nanowires are smaller than the measured physical dimensions. An interesting comparison for these samples is of the sample dimensions to the superconducting coherence length in NbN in the dirty limit, expected to be a few nanometres \cite{jesudasan}. We have observed behaviour characteristic of CQPS in samples that ostensibly have a width and a thickness of a factor up to $\sim$10-times greater than $\xi$. At first sight, this is rather surprising. It is not in fact a strict requirement that a nanowire have $w,d<\xi$ in order for QPS to be observed. Rather, since the smallest volume in which superconductivity is required to fluctuate to enable a $2\pi$ slip in the phase difference between the two sides is $wd\xi$, the energy barrier for such a fluctuation, the~superconducting condensation energy of this volume, is proportional to the volume fluctuating. Where~either or both of the cross-sectional dimensions exceed $\xi$, this therefore leads to an additional factor $\xi/w$ and/or $\xi/d$ multiplying the numerical factor $b$ in Equation \eqref{eq}, leading to an expected exponential suppression of $E_\textrm{S}$ by factors $d/\xi$ and $w/\xi$ whenever either of these is greater than 1. Our~nanowires are significantly wider and thicker than the coherence length, and yet we observe behaviour characteristic of CQPS. There are a number of possible explanations for this. It is conceivable, though seems unlikely, that the QPS energy scale for the material is extremely large, meaning that even when exponentially suppressed, it is still significant, and that nanowires with dimensions $\approx \xi$ would have such a large $E_\textrm{S}$ that measurements would show zero conductance for any practical value of voltage bias. Another, perhaps more likely, possibility is that the relevant volume fluctuating is smaller than the value $wd\xi$ determined from the measured physical dimensions $w$ and $d$. The cross-sectional area of the nanowire carrying the supercurrent could be smaller than the physical dimensions as a result of a dead layer at the interface with the substrate and the external surfaces as a result of oxidation after deposition. It is also possible that inhomogeneities in the material could give a distribution of cross-sectional areas along the length of the nanowire. In that case, in view of the exponential dependence of the QPS energy on the cross-sectional area, the properties would be dominated by the region or regions of the nanowire with the smallest cross-sectional area \cite{vanevic}. A distribution in cross-sectional areas could arise either as a result of variation in the physical dimensions along the length of the nanowire (see Section \ref{fabchall}) or as a result of more microscopic inhomogeneity such as is expected to be induced close to the superconductor--insulator transition \cite{feigelman} and that might lead to a percolating pathway through the nanowire.

While the nanowire length has not been a major focus of this study, we are nonetheless able to make some inferences in relation to nanowire length. The nanowires investigated in this study were $\sim$1--10 $\upmu$m long, similar to the lengths employed in References \citenum{hongisto,websterfenton}, but longer than the nanowires employed in References \citenum{astafiev, peltonen, degraaf}, which were in the range of 30--750 nm. Two previous works have argued that collective excitations in longer nanowires might suppress CQPS effects \cite{kerman,cedergren}. The~present observations of critical voltage features in long nanowires, along with the already-reported observations \cite{hongisto,websterfenton}, provide basic empirical experimental evidence that CQPS effects are still relevant in long nanowires, although it remains an open question whether these long nanowires should actually rather be viewed as a series of shorter nanowires because of inhomogeneity, as discussed above.

Nanowire NbN80/1, in which the substantial critical voltage $\approx$5 mV has been observed, was one of only two nanowires reported here in which series inductance and resistance were included. Nanowire NbN81/2, deposited in the same experimental run and nominally having a very similar width, as well as also similarly incorporating series inductance and resistance, shows no $V_\textrm{c}$ feature at all. It appears likely that a composition closer to the superconductor--insulator transition, as indicated by the $R(T)$ dependence, is a critical factor in determining the substantial $V_\textrm{c}$ for nanowire NbN80/1; the series inductance and resistance elements may also be important, but are clearly not sufficient. The~different compositions in the two co-deposited samples may indicate that the thickness of nanowire NbN80/1 is smaller, as a result of thickness variations during deposition or more likely due to over-etching. It is clear that the observations of these two nanowires indicate variability in nanowire properties and highlight an outstanding challenge in reproducibly obtaining nanowires exhibiting strong QPS.

\subsection{Fabrication Challenges} \label{fabchall}
The different fabrication methods we have utilised have different characteristic features and different advantages and disadvantages. Using both EBL and Ne-FIB, we have obtained nanowires with widths $<20$ nm. Nanowire widths of 15--20 nm are routine to generate using the EBL negative-resist process using hydrogen silsesquioxane (HSQ) as the resist. We have obtained nanowire widths of \mbox{$\approx$20 nm} via the cut-out technique, using polymethyl methacrylate (PMMA) as the resist. Ne-FIB is able to generate nanowires with widths below 25 nm, although fabrication becomes more challenging for the narrowest nanowires. Although the results we report showing a critical voltage feature have been measured in samples fabricated using the negative resist technique with HSQ, we believe that the ultimate widths obtainable by the other two techniques are small enough that they are also suitable for use for fabrication of nanowires for use as CQPS elements.

The processes determining the morphology and roughness of the nanowires are different for nanowires fabricated via EBL and via Ne-FIB. For EBL-fabricated (negative resist or positive-resist cut-out) nanowires, the sidewall morphology is affected by the sidewall of the resist mask. The edges of a resist mask with non-vertical sidewalls will be etched through prior to completion of etching, leading~to transfer of the resist mask profile into the NbN nanowire. Where nonuniformities in thickness of the resist are present, these might also be transferred into the nanowire profile --- an issue that is more relevant for PMMA resist as the polymer molecule sizes are $\approx$20 nm, with inhomogeneities in the resist at shorter length scales limiting the uniformity of the feature definition at those scales. The~chemistry of the reactive ion etch process and the isotropic element of its chemical action also affect the morphology and roughness of the NbN material remaining after RIE. In particular, since different components of the nanowire, such as oxides, etch at different rates, the RIE processing will tend to accentuate inhomogeneities present in the deposited film as the nanowire is defined. 

Etching by Ne-FIB, on the other hand, is directionally highly anisotropic, a feature that is beneficial for well-defined vertical sidewalls. Of course, this process is not, in practice, perfect. The~ion beam has a Gaussian beam profile rather than being perfectly confined; however, the limiting factor for sidewall definition is actually scattering once the beam enters the material \cite{Tan2010}, and this leads to damage to and/or milling of regions beyond the diameter of the incident beam. We take care during Ne-FIB processing to minimise the ion dose supplied to parts of the sample other than those we are milling, in order to minimise damage to other regions of the sample. Some initial imaging is required in order to locate the region to be milled. Our FIB instrument also has the ability to make use of helium ions, which, being lighter atoms, cause even less damage. We typically initially locate the region to be milled while imaging with helium, before switching to neon for the milling. In practice, obtaining very fine features relies on the drift due to charging and mechanical effects during the time required for milling being small compared to the resolution required. Sample drift due to charging may be minimised by prior wire-bonding of the sample bond pads to ground, and charging during imaging may be further reduced by the use of an electron flood gun, which allows neutralisation of charging of the substrate as a result of the ion-beam bombardment. 

For samples fabricated using a negative-resist mask, with HSQ, several experimental challenges have informed the parameters we use. Firstly, the thickness of the HSQ layer should be chosen to satisfy two competing requirements. The HSQ layer is etched along with the NbN during the RIE stage, albeit at a slower rate, and therefore must be sufficiently thick to protect the NbN nanowire underneath it while the unwanted areas of NbN are removed. The maximum resist thickness is set by aspect-ratio considerations: an HSQ feature that, after EBL patterning, has a high aspect ratio, {i.e.}, its thickness is much greater than its width, is mechanically unstable and may fall over during development or drying. The ideal thickness for the HSQ layer would be such that it is just removed at the end of the RIE stage. 

Adhesion during patterning is an issue that we have tackled. While good adhesion of HSQ to the underlying NbN film for features with both in-plane dimensions $\gtrapprox 1$ $\upmu$m is found, in some samples the narrow nanowire sections of the HSQ resist-mask apparently can float off the NbN film during development. While even long nanowire sections of the resist-mask may sometimes land intact on the substrate following development \cite{fentoneucas}, this is an undesirable feature for a fabrication process, and we have addressed the problem by introducing ``anchoring'' when using long sections of NbN, whether nanowire or inductive line, that is, introducing periodic anchors into the inductive line which have both in-plane dimensions $>1$ $\upmu$m. We have observed an issue of kinks developing in long nanowires (as seen to a small extent in Figure \ref{HeFIB}a), a factor that favours the fabrication of sections of nanowire shorter than several $\upmu$m in length.
We expect that both the aspect-ratio and adhesion issues would also not be an issue in short-enough nanowires, since such structures are more mechanically stable against lateral forces applied to the top of the nanowire furthest from its ends.

\subsection{IV Measurement Considerations} \label{meastdisc}
$I(V)$ measurements should be made using careful filtering \cite{arutyunov}. In the comparisons we have made of the temperature variation of the resistance with and without the use of filters (not shown here), we~have observed, when measuring without filtering, a resistive tail below $T_\textrm{c}$ that resembles the resistive tails frequently observed in nanowires and successfully fitted to models of thermally activated or quantum phase-slips. However, when the measurement is repeated with proper filtering, no tail is observed. This implies that the observed resistive tail was an experimental artefact induced by noise being conducted down the measurement lines and shows the importance of careful filtering of experimental lines.

Another highly relevant issue in measurements of nanowires undergoing coherent quantum phase-slips is the presence of long time-constants before the steady state response is reached following a change in bias \cite{arutyunovjsnm}. As mentioned above, it is important to embed the nanowire in a high-impedance environment with $R>R_\textrm{Q}$, and an effect associated with this is that the combination of this high impedance with the cable, filter and parasitic capacitances acts as a low-pass filter removing high-frequency noise from the measurement lines. However, a superconducting nanowire undergoing CQPS can behave as an impedance $\gg R_\textrm{Q}$, and the combination of this impedance and the resistance of series thin-film resistance with cable, filter and/or parasitic capacitances leads to $RC$ time constants for the capacitances to charge up to an equilibrium state, during which time a fraction of the current registered in the experimental current measurement is actually flowing into these capacitances rather than through the sample. The time constants can typically be 1--10 s, and in extreme cases above 100~s, in which case measurements become impractical. This should be a consideration in the design of filters and cabling for measurements on these nanowire systems; in particular, capacitances included in filters should not be made too high-value.

A further consideration in circuit design relates to mitigating self-heating in the devices. Since~coupling between electron and phonon systems is weak at mK temperatures, such devices are prone to overheating when dissipation is present. In order to minimise such heating, thin-film resistors should be designed to have as large a volume as possible to achieve the required resistance value \cite{websterfenton}, since~maximising the volume maximises the interaction between the electron and phonon systems.

\section{Materials and Methods}
\subsection{Film Deposition}
NbN films are deposited at room temperature on sapphire or silicon substrates using reactive DC magnetron sputtering from a Nb target in a nitrogen-containing atmosphere. Using a pressure of $5\times 10^{-3}$ mbar with a 1:1 flow of Ar and N$_2$ gas, 150-W sputter power gives a deposition rate of around 10 nm/min. 

\subsection{Nanowire Definition}
In this section, we present the experimental details of the three methods we have used to fabricate nanowires. These are shown schematically in Figure~\ref{fabtech}, and Figure~\ref{HeFIB} shows images of one sample fabricated by each of the methods, collected using a helium focussed ion-beam or scanning electron microscope.

Our first technique for defining nanowires employs a negative resist to define a mask. We use hydrogen silsesquioxane (HSQ) resist, diluted to 1\% HSQ in methyl isobutyl ketone (MIBK) and spun at 2000 rpm to give a 35-nm HSQ layer. To achieve nanowires with widths as low as 15--20 nm, we used a 10-kV electron beam to write single-pixel lines in the resist using a line dose set at some multiple of a default value of 1280 pC/cm, followed by development for one minute in MF-26A~developer. The nanowire is then formed using reactive ion etching (RIE) at 100 W and 100 mTorr using flows of 35 sccm of CHF$_3$ and 14 sccm of SF$_6$ for approximately 120 s in 20-s bursts to mitigate potential overheating during etching. Since removal of exposed HSQ requires the use of HF, we do not remove any HSQ remaining after RIE.

For the second technique we have used for defining nanowires via EBL, we utilise polymethyl methacrylate (PMMA), a positive-tone resist, and follow a ``cut-out'' strategy \cite{constantino} to expose the edges of the nanowires to be removed, followed by RIE, using the same recipe as above. This~method means that HSQ resist is not required, and the PMMA resist may be removed in acetone when etching is complete. Figure~\ref{HeFIB}b shows an image of a nanowire prepared using this method. The~`cut-out' lines are typically $\sim$20~nm wide. The data shown in Figure~\ref{NCnwIVs} were obtained on nanowires prepared in this way. An advantage of this technique is that the ultimate width obtainable by the technique may be made smaller than the narrowest line obtainable by patterning the PMMA resist, since the linewidth is determined by the difference between the separation of the centres of the `cut-out' lines and the width of the cut-out line, which may be chosen when the pattern is exposed by EBL. Note also that over-exposure in this geometry leads to a reduction in the width of the nanowire obtained, rather than an increase in the width as would be obtained when patterning using a negative resist. As for any EBL-based process in which ultimate resolution is sought, the ultimate linewidth is obtained following suitable dose tests and relies on the stability of results obtained by the EBL and subsequent development and etching.

The third fabrication technique we have used is based on neon focussed ion-beam milling. This does not rely on either an e-beam resist mask or RIE when defining the nanowire and allows milling with 5-nm resolution. Although some damage is inevitable in focussed ion-beam milling, since neon ions are inert (unlike the more commonly-used gallium), poisoning from implanted ions is expected to be avoided when milling using neon. We typically use a Ne ion beam accelerated to 15 kV, providing a beam current of $\sim$2 pA, to supply a dose of 0.5--1 nC/$\upmu$m$^2$ to remove material from the NbN film to define the nanowire. Because of the relatively slow speed of milling, it is not practicable to fabricate the whole structure by FIB, so the deposited NbN film is first coated with PMMA and patterned by EBL, then etched by RIE to define the coarse features of the structure, in a similar way to the ``cut-out'' strategy. A wider nanowire (with a width of 300 nm) is patterned in this step, and this nanowire is then milled using the Ne-FIB to remove material in order to define a nanowire. In previously published work, we have also used this technique successfully to fabricate NbN nanowires as elements within superconducting co-planar waveguide resonators \cite{burnetteucas, burnett}.
 
\subsection{Other Circuit Components}
For some nanowire samples, we fabricate additional components in series with the narrow nanowire. Wider sections of NbN, deposited and patterned in the same steps as the narrow sections of nanowire, may, as a result of the high kinetic inductance of the NbN, conveniently be used to provide series inductance for the circuit. The wires are narrow in order to provide the maximum amount of inductance in the shortest length, but are made wide enough that they do not themselves experience significant QPS effects. Typically, we fabricate these inductor wires with widths of $\approx$200 nm.

To provide series resistance for the nanowire circuits, we use electron beam lithography, reactive DC magnetron sputtering and lift-off to produce thin-film chromium oxide resistors with typical sheet resistance $\sim 1$ k$\Omega$, in a process we have described in detail elsewhere \cite{nash}. We pattern gold interconnects between resistors and NbN components, wiring and bond pads using EBL, DC magnetron sputtering and lift-off, using a Ti or Cr adhesion layer and an argon-ion mill clean step {in situ} immediately prior to gold deposition. Prior to EBL to define resistors or gold wiring, a gold layer with a thickness of $\sim$10 nm is sputtered on top of the resist layer. This conductive gold layer prevents charging of the insulating sapphire substrate and is removed by a KI/I$_2$/H$_2$O wet-etch dip immediately prior to each development stage \cite{fentoneucas}.

\section{Conclusions}
We have presented three different techniques for fabricating superconducting nanowires suitable for use as coherent quantum phase-slip elements. We have obtained nanowires with a width $<30$ nm using all three techniques, and all three appear promising for fabricating nanowires for use as CQPS elements. We have characterised the effect of reducing both thickness and width dimensions on the superconducting properties of NbN and have studied several NbN nanowires with a composition approaching the superconductor--insulator transition. We have observed a variety of behaviours in the nanowires; these behaviours include standard superconducting behaviour as would characterise wide and thick superconducting tracks and phase-slip centres, as well as current-voltage characteristics as characterise incoherent QPS and zero conductance below a critical voltage $V_\textrm{c}$ as characterises coherent QPS. We have observed critical voltages $V_\textrm{c}$ up to 5 mV, an order of magnitude larger than in previous reports. We have confirmed both that the $V_\textrm{c}$ feature is accompanied by a critical-current feature at higher bias and that the $V_\textrm{c}$ feature develops over the same temperature range as the superconductivity, indicating that the $V_\textrm{c}$ feature has a superconducting origin, as expected for a feature associated with CQPS. We have observed $V_\textrm{c}$ features in nanowires with lengths $\sim$10 $\upmu$m, suggesting that it is not essential that CQPS elements are constituted from short nanowires. We have observed a $V_\textrm{c}$ feature in nanowires with cross-sectional dimensions exceeding the superconducting coherence length, suggesting the importance of materials inhomogeneity in determining the properties of nanowires. Our results also indicate the importance of material composition in the properties of superconducting nanowires with dimensions $\lesssim$20 nm, with the use of filtered measurement lines and a high-impedance environment being preferable to promote CQPS behaviour. The success of the presented fabrication technologies and observation of enhanced critical voltage are promising for applications of superconducting nanowires in coherent quantum phase-slip applications, and the variety of behaviour observed in similar nanowires also points to a remaining challenge of reproducibly controlling the properties of fabricated nanowires.

\acknowledgments{
This research was funded by the U.K. Engineering and Physical Sciences Research Council Grant Numbers EP/
J017329/1 (J.C.F., M.S.A.) and EP/H005544/1 (P.A.W.). O.W.K. acknowledges funding from Carl Zeiss Semiconductor Manufacturing Technology. The authors are grateful for the technical assistance of Suguo Huo.}


\begin{thebibliography}{999}

\bibitem[Doyle \em{et~al.}(2010)Doyle, Mauskopf, Zhang, Monfardini, Swenson,
Baselmans, Yates, and Roesch]{doylereview}
Doyle, S.; Mauskopf, P.; Zhang, J.; Monfardini, A.; Swenson, L.; Baselmans, J.;
Yates, S.; Roesch, M.
\newblock A review of the lumped element kinetic inductance detector.
\newblock {\em Proc. SPIE} {\bf 2010}, {\em 7741},
\newblock
doi:10.1117/12.857341.

\bibitem[Hadfield(2009)]{hadfieldreview}
Hadfield, R.
\newblock Single-photon detectors for optical quantum information applications.
\newblock {\em Nat. Photonics} {\bf 2009}, {\em 3},~696,
\newblock
doi:{10.1038/nphoton.2009.230}.

\bibitem[Mooij and Nazarov(2006)]{mooijnazarov}
Mooij, J.; Nazarov, Y.
\newblock Superconducting nanowires as quantum phase-slip junctions.
\newblock {\em Nat. Phys.} {\bf 2006}, {\em 2},~169--172,
\newblock
doi:{10.1038/nphys234}.

\bibitem[Hriscu and Nazarov(2011)]{hriscu}
Hriscu, A.; Nazarov, Y.
\newblock Coulomb blockade due to quantum phase-slips illustrated with devices.
\newblock {\em Phys. Rev. B} {\bf 2011}, {\em 83},~174511,
\newblock
doi:{{{10.1103/PhysRevB.83.174511}}}.

\bibitem[Tinkham(1996)]{tinkham}
Tinkham, M.
\newblock {\em Introduction to Superconductivity}, 2nd ed.; McGraw--Hill: New York, NY, USA,
1996.

\bibitem[Chandrasekhar and Webb(1994)]{chandrasekhar}
Chandrasekhar, V.; Webb, R.
\newblock Single electron charging effects in high-resistance In$_2$O$_{3-x}$
Wires.
\newblock {\em J. Low Temp.~Phys.} {\bf 1994}, {\em 97},~9,
\newblock
doi:{{{10.1007/BF00752978}}}.

\bibitem[Haviland \em{et~al.}(1991)Haviland, Kuzmin, Delsing, and
Claeson]{haviland}
Haviland, D.; Kuzmin, L.; Delsing, P.; Claeson, T.
\newblock Observation of the Coulomb blockade of Cooper pair tunnelling in
single Josephson junctions.
\newblock {\em Europhys. Lett.} {\bf 1991}, {\em 16},~103,
\newblock
doi:{{{10.1209/0295-5075/16/1/018}}}.

\bibitem[Astafiev \em{et~al.}(2012)Astafiev, Ioffe, Kafanov, Pashkin,
Arutyunov, Shahar, Cohen, and Tsai]{astafiev}
Astafiev, O.; Ioffe, L.; Kafanov, S.; Pashkin, Y.A.; Arutyunov, K.Y.; Shahar,
D.; Cohen, O.; Tsai, J.
\newblock Coherent quantum phase-slip.
\newblock {\em Nature} {\bf 2012}, {\em 484},~355--358,
\newblock
doi:{{{10.1038/nature10930}}}.

\bibitem[Webster \em{et~al.}(2013)Webster, Fenton, Hongisto, Giblin, Zorin, and
Warburton]{websterfenton}
Webster, C.; Fenton, J.; Hongisto, T.; Giblin, S.; Zorin, A.; Warburton, P.
\newblock NbSi nanowire quantum phase-slip circuits: DC supercurrent blockade,
microwave measurements, and thermal analysis.
\newblock {\em Phys. Rev. B} {\bf 2013}, {\em 87},~144510,
\newblock
doi:{{{10.1103/PhysRevB.87.144510}}}.

\bibitem[Hongisto and Zorin(2012)]{hongisto}
Hongisto, T.; Zorin, A.
\newblock Single-charge transistor based on the charge-phase duality of a
superconducting nanowire circuit.
\newblock {\em Phys. Rev. Lett.} {\bf 2012}, {\em 108},~097001,
\newblock
doi:{{{10.1103/PhysRevLett.108.097001}}}.

\bibitem[Peltonen \em{et~al.}(2013)Peltonen, Astafiev, Korneeva, Voronov,
Korneev, Charaev, Semenov, Golt'sman, Ioffe, Klapwijk, et~al.]{peltonen}
Peltonen, J.; Astafiev, O.; Korneeva, Y.P.; Voronov, B.; Korneev, A.; Charaev,
I.; Semenov, A.; Golt'sman, G.; Ioffe, L.; Klapwijk, T.; et al.
\newblock Coherent flux tunneling through NbN nanowires.
\newblock {\em Phys. Rev. B} {\bf 2013}, {\em 88},~220506,
\newblock
doi:{{{10.1103/PhysRevB.88.220506}}}.

\bibitem[de~Graaf \em{et~al.}(2018)de~Graaf, Skacel, H\"{o}nigl-Decrinis,
Shaikhaidarov, Rotzinger, Linzen, Ziegler, H\"{u}bner, Meyer, Antonov,
Il'ichev, Ustinov, Tzalenchuk, and Astafiev]{degraaf}
De~Graaf, S.; Skacel, S.; H\"{o}nigl-Decrinis, T.; Shaikhaidarov, R.;
Rotzinger, H.; Linzen, S.; Ziegler, M.; H\"{u}bner, U.; Meyer, H.; Antonov, V.;
et al.
\newblock Charge quantum interference device.
\newblock {\em Nat. Phys.} {\bf 2018},
\newblock
doi:{{{10.1038/s41567-018-0097-9}}}.

\bibitem[Arutyunov \em{et~al.}(2016)Arutyunov, Lehtinen, and
Rantala]{arutyunovjsnm}
Arutyunov, K.; Lehtinen, J.; Rantala, T.
\newblock The quantum phase-slip phenomenon in superconducting nanowires with
high-impedance environment.
\newblock {\em J. Supercond. Nov. Magn.} {\bf 2016},
{\em 29},~569,
\newblock
doi:{{{10.1007/s10948-015-3298-9}}}.

\bibitem[Kafanov and Chtchelkatchev(2013)]{kafanov}
Kafanov, S.; Chtchelkatchev, N.
\newblock Single flux transistor: The controllable interplay of coherent
quantum phase-slip and flux quantization.
\newblock {\em J. Appl. Phys.} {\bf 2013}, {\em 114},~073907,
\newblock
doi:{{{10.1063/1.4818706}}}.

\bibitem[Goldman and Markovic(1998)]{goldman}
Goldman, A.; Markovic, N.
\newblock Superconductor--insulator transitions in the two-dimensional limit.
\newblock {\em Phys. Today} {\bf 1998},~39,
\newblock
doi:{{{10.1063/1.882069}}}.

\bibitem[Paalanen \em{et~al.}(1992)Paalanen, Hebard, and Ruel]{paalanen}
Paalanen, M.; Hebard, A.; Ruel, R.
\newblock Low-temperature insulating phases of uniformly disordered
two-dimensional superconductors.
\newblock {\em Phys. Rev. Lett.} {\bf 1992}, {\em 69},~1604,
\newblock
doi:{{{10.1103/PhysRevLett.69.1604}}}.

\bibitem[Bollinger \em{et~al.}(2008)Bollinger, Dinsmore, Rogachev, and
Bezryadin]{bollinger}
Bollinger, A.; Dinsmore, R.; Rogachev, A.; Bezryadin, A.
\newblock Determination of the superconductor--insulator phase diagram for
one-dimensional wires.
\newblock {\em Phys. Rev. Lett.} {\bf 2008}, {\em 101},~227003,
\newblock
doi:{{{10.1103/PhysRevLett.101.227003}}}.

\bibitem[Lau \em{et~al.}(2001)Lau, Markovic, Bocrath, Bezryadin, and
Tinkham]{lau}
Lau, C.; Markovic, N.; Bocrath, M.; Bezryadin, A.; Tinkham, M.
\newblock Quantum phase-slips in superconducting nanowires.
\newblock {\em Phys. Rev. Lett.} {\bf 2001}, {\em 87},~217003,
\newblock
doi:{{{10.1103/PhysRevLett.87.217003}}}.

\bibitem[Arutyunov \em{et~al.}(2008)Arutyunov, Golubev, and Zaikin]{arutyunov}
Arutyunov, K.; Golubev, D.; Zaikin, A.
\newblock Superconductivity in one dimension.
\newblock {\em Phys. Rep.} {\bf 2008}, {\em 464},~1,
\newblock
doi:{{{10.1016/j.physrep.2008.04.009}}}.

\bibitem[Altomare \em{et~al.}(2006)Altomare, Chang, Melloch, Hong, and
Tu]{altomare}
Altomare, F.; Chang, A.; Melloch, M.; Hong, Y.; Tu, C.
\newblock Evidence for macroscopic quantum tunneling of phase-slips in long
one-dimensional superconducting Al wires.
\newblock {\em Phys. Rev. Lett.} {\bf 2006}, {\em 97},~017001,
\newblock
doi:{{{10.1103/PhysRevLett.97.017001}}}.

\bibitem[Makise \em{et~al.}(2016)Makise, Terai, Tominari, Tanaka, and
Shinozaki]{makise}
Makise, K.; Terai, H.; Tominari, Y.; Tanaka, S.; Shinozaki, B.
\newblock Duality picture of superconductor--insulator transitions on
superconducting nanowire.
\newblock {\em Sci. Rep.} {\bf 2016}, {\em 6},~27001,
\newblock
doi:{{{10.1038/srep27001}}}.

\bibitem[Fenton \em{et~al.}(2011)Fenton, Webster, and Warburton]{fentonjpcs}
Fenton, J.; Webster, C.; Warburton, P.
\newblock Materials for superconducting nanowires for quantum phase-slip
devices.
\newblock {\em J. Phys. Conf. Ser.} {\bf 2011}, {\em
286},~012024,
\newblock
doi:{{{10.1088/1742-6596/286/1/012024}}}.

\bibitem[Mooij \em{et~al.}(2015)Mooij, Sch\"{o}n, Shnirman, Fuse, Harmans,
Rotzinger, and Verbruggen]{mooijnjp}
Mooij, J.; Sch\"{o}n, G.; Shnirman, A.; Fuse, T.; Harmans, C.; Rotzinger, H.;
Verbruggen, A.
\newblock Superconductor–insulator transition in nanowires and nanowire
arrays.
\newblock {\em New J. Phys.} {\bf 2015}, {\em 17},~033006,
\newblock
doi:{{{10.1088/1367-2630/17/3/033006}}}.

\bibitem[Kautz(1981)]{kautz}
Kautz, R.
\newblock Chaotic states of rf‐biased Josephson junctions.
\newblock {\em J. Appl. Phys.} {\bf 1981}, {\em 52},~641,
\newblock
doi:{{{10.1063/1.328566}}}.

\bibitem[Kim \em{et~al.}(2012)Kim, Jamali, and Rogachev]{kim}
Kim, H.; Jamali, S.; Rogachev, A.
\newblock Superconductor--insulator transition in long MoGe nanowires.
\newblock {\em \mbox{Phys. Rev. Lett}.} {\bf 2012}, {\em 109},~027002,
\newblock
doi:{{{10.1103/PhysRevLett.109.027002}}}.

\bibitem[Arutyunov \em{et~al.}(2012)Arutyunov, Hongisto, Lehtinen, Leino, and
Vasiliev]{arutyunovscirep}
Arutyunov, K.; Hongisto, T.; Lehtinen, J.; Leino, L.; Vasiliev, A.
\newblock Quantum phase-slip phenomenon in ultra-narrow superconducting
nanorings.
\newblock {\em Sci. Rep.} {\bf 2012}, {\em 2},~293,
\newblock
doi:{{{10.1038/srep00293}}}.

\bibitem[Lehtinen \em{et~al.}(2017)Lehtinen, Kemppinen, Mykk\"{a}nen, Prunnila,
and Manninen]{manninen}
Lehtinen, J.; Kemppinen, A.; Mykk\"{a}nen, E.; Prunnila, M.; Manninen, A.
\newblock Superconducting MoSi nanowires.
\newblock {\em Supercond. Sci. Technol.} {\bf 2017}, {\em
31},~015002,
\newblock
doi:{{{10.1088/1361-6668/aa954b}}}.

\bibitem[Nash \em{et~al.}(2014)Nash, Fenton, Constantino, and Warburton]{nash}
Nash, C.; Fenton, J.; Constantino, N.; Warburton, P.
\newblock Compact chromium oxide thin film resistors for use in nanoscale
quantum circuits.
\newblock {\em J. Appl. Phys.} {\bf 2014}, {\em 116},~224501,
\newblock
doi:{{{10.1063/1.4901933}}}.

\bibitem[Feigel'man \em{et~al.}(2010)Feigel'man, Ioffe, Kravtsov, and
Cuevas]{feigelman}
Feigel'man, M.; Ioffe, L.; Kravtsov, V.; Cuevas, E.
\newblock Fractal superconductivity near localization threshold.
\newblock {\em Ann.~Phys.} {\bf 2010}, {\em 325},~1390,
\newblock
doi:{{{10.1016/j.aop.2010.04.001}}}.

\bibitem[Constantino(2016)]{constantino}
Constantino, N.G.N.
\newblock Disorder in Superconductors in Reduced Dimensions.
\newblock Ph.D. Thesis, University College London, Gower Street, London,
UK, 2016.

\bibitem[Finkel'stein(1994)]{finkelstein}
Finkel'stein, A.
\newblock Suppression of superconductivity in homogeneously disordered systems.
\newblock {\em Phys. B Condens.~Matter} {\bf 1994}, {\em 197},~636,
\newblock
doi:{{{10.1016/0921-4526(94)90267-4}}}.

\bibitem[Ivry \em{et~al.}(2014)Ivry, Kim, Dane, De~Fazio, McCaughan, Sunter,
Zhao, and Berggren]{ivry}
Ivry, Y.; Kim, C.; Dane, A.; De~Fazio, D.; McCaughan, A.; Sunter, K.; Zhao, Q.;
Berggren, K.
\newblock Universal scaling of the critical temperature for thin films near the
superconducting-to-insulating transition.
\newblock {\em Phys. Rev. B} {\bf 2014}, {\em 90},~214515,
\newblock
doi:{{{10.1103/PhysRevB.90.214515}}}.

\bibitem[Crauste \em{et~al.}(2013)Crauste, Gentils, Cou\"{e}do, Dolgorouky,
Berg\'{e}, Collin, Marrache-Kikuchi, and Dumoulin]{crauste}
Crauste, O.; Gentils, A.; Cou\"{e}do, F.; Dolgorouky, Y.; Berg\'{e}, L.;
Collin, S.; Marrache-Kikuchi, C.; Dumoulin, L.
\newblock Effect of annealing on the superconducting properties of
a-Nb$_x$Si$_{1-x}$ thin films.
\newblock {\em Phys. Rev. B} {\bf 2013}, {\em 87},~144514,
\newblock
doi:{{{10.1103/PhysRevB.87.144514}}}.

\bibitem[Fenton and Burnett(2016)]{fentoneucas}
Fenton, J.; Burnett, J.
\newblock Superconducting NbN Nanowires and Coherent Quantum Phase-Slips in DC
Transport.
\newblock {\em IEEE Trans. Appl. Supercond.} {\bf 2016}, {\em
26},~2200505,
\newblock
doi:{{{10.1109/TASC.2016.2531005}}}.

\bibitem[Jesudasan \em{et~al.}(2011)Jesudasan, Mondal, Chand, Kamlapure, Kumar,
Saraswat, Bagwe, Tripathi, and Raychaudhuri]{jesudasan}
Jesudasan, J.; Mondal, M.; Chand, M.; Kamlapure, A.; Kumar, S.; Saraswat, G.;
Bagwe, V.; Tripathi, V.; Raychaudhuri, P.
\newblock Upper Critical Field and Coherence Length of Homogenously Disordered
Epitaxial 3‐Dimensional NbN Films.
\newblock {\em AIP Conf. Proc.} {\bf 2011}, {\em 923},~1349,
\newblock
doi:{{{10.1063/1.3606160}}}.

\bibitem[Vanevic and Nazarov(2012)]{vanevic}
Vanevic, M.; Nazarov, Y.
\newblock Quantum phase-slips in superconducting wires with weak
inhomogeneities.
\newblock {\em \mbox{Phys. Rev. Lett}.} {\bf 2012}, {\em 108},~187002,
\newblock
doi:{{{10.1103/PhysRevLett.108.187002}}}.

\bibitem[Kerman(2013)]{kerman}
Kerman, A.
\newblock Flux–charge duality and topological quantum phase fluctuations in
quasi-one-dimensional superconductors.
\newblock {\em New J. Phys.} {\bf 2013}, {\em 15},~105017,
\newblock
doi:{{{10.1088/1367-2630/15/10/105017}}}.

\bibitem[Cedergren \em{et~al.}(2017)Cedergren, Ackroyd, Kafanov, Vogt,
Shnirman, and Duty]{cedergren}
Cedergren, K.; Ackroyd, R.; Kafanov, S.; Vogt, N.; Shnirman, A.; Duty, T.
\newblock Insulating Josephson Junction chains as pinned Luttinger liquids.
\newblock {\em Phys. Rev. Lett.} {\bf 2017}, {\em 119},~167701,
\newblock
doi:{{{10.1103/PhysRevLett.119.167701}}}.

\bibitem[Tan \em{et~al.}(2010)Tan, Livengood, Shima, Notte, and McVey]{Tan2010}
Tan, S.; Livengood, R.; Shima, D.; Notte, J.; McVey, S.
\newblock {Gas field ion source and liquid metal ion source charged particle
material interaction study for semiconductor nanomachining applications}.
\newblock {\em J. Vac. Sci. Technol. B Microelectron.
 Nanometer Struct.} {\bf 2010}, {\em 28},~C6F15,
\newblock
doi:{{{10.1116/1.3511509}}}.

\bibitem[Burnett and Fenton(2016)]{burnetteucas}
Burnett, J.; Fenton, J.
\newblock Embedding NbN nanowires into quantum circuits with a neon focused ion
beam.
\newblock {\em IEEE~Trans. Appl. Supercond.} {\bf 2016}, {\em
26},~1700104,
\newblock
doi:{{{10.1109/TASC.2016.2525988}}}.

\bibitem[Burnett \em{et~al.}(2017)Burnett, Sagar, Kennedy, Warburton, and
Fenton]{burnett}
Burnett, J.; Sagar, J.; Kennedy, O.; Warburton, P.; Fenton, J.
\newblock Low-loss superconducting nanowire circuits using a neon focused ion
beam.
\newblock {\em Phys. Rev. Appl.} {\bf 2017}, {\em 8},~014039,
\newblock
doi:{{{10.1103/PhysRevApplied.8.014039}}}.


\end{thebibliography}
\end{document}